\documentclass{jov}
\usepackage[utf8]{inputenc} 
\usepackage{graphicx}

\begin{document}
\title{Matching visual induction effects on screens of \\ different size}
\abstract{\textit{This work is dedicated to the memory of Jihyun Yeonan-Kim.}
\\
\\In the film industry, the same movie is expected to be watched on displays of vastly different sizes, from cinema screens to mobile phones.
But visual induction, the perceptual phenomenon by which the appearance of a scene region is affected by its surroundings, will be different for the same image shown on two displays of different dimensions. This presents a practical challenge for the preservation of the artistic intentions of filmmakers, as it can lead to shifts in image appearance between viewing destinations.
In this work we show that a neural field model based on the efficient representation principle is able to predict induction effects, and how by regularizing its associated energy functional the model is still able to represent induction but is now invertible.
From this we propose a method to pre-process an image in a screen-size dependent way so that its perception, in terms of visual induction, may remain constant across displays of different size. The potential of the method is demonstrated through psychophysical experiments on synthetic images and qualitative examples on natural images.}

\author{Canham}{Trevor}
 {Department of Information and Communication Technologies}
 {Universitat Pompeu Fabra, Barcelona, Spain}
 {http://}{trevor.canham@upf.edu}
 \author{Vazquez-Corral}{Javier}
 {Department of Information and Communication Technologies}
 {Universitat Pompeu Fabra, Barcelona, Spain}
{http://www.jvazquez-corral-net}{javier.vazquez@upf.edu}
\author{Mathieu}{Elise}
{Grenoble INP - Phelma}
 {France}
 {http://}{Elise.Mathieu1@grenoble-inp.org}
\author{Bertalm{\'\i}o}{Marcelo}
{Department of Information and Communication Technologies}
 {Universitat Pompeu Fabra, Barcelona, Spain}
 {http://}{marcelo.bertalmio@upf.edu}

\maketitle

\section{1. Introduction}

\begin{figure}[ht!]
    \centering
    {\includegraphics[width=0.3\textwidth]{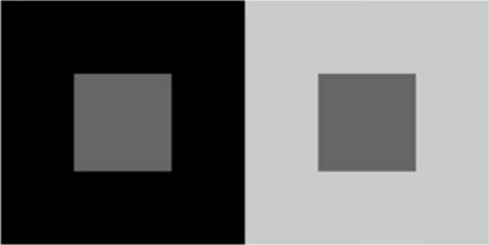} }%
    \qquad
    {\includegraphics[width=0.3\textwidth]{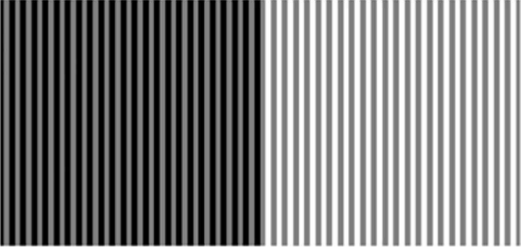} }%
    \qquad
    {\includegraphics[width=0.3\textwidth]{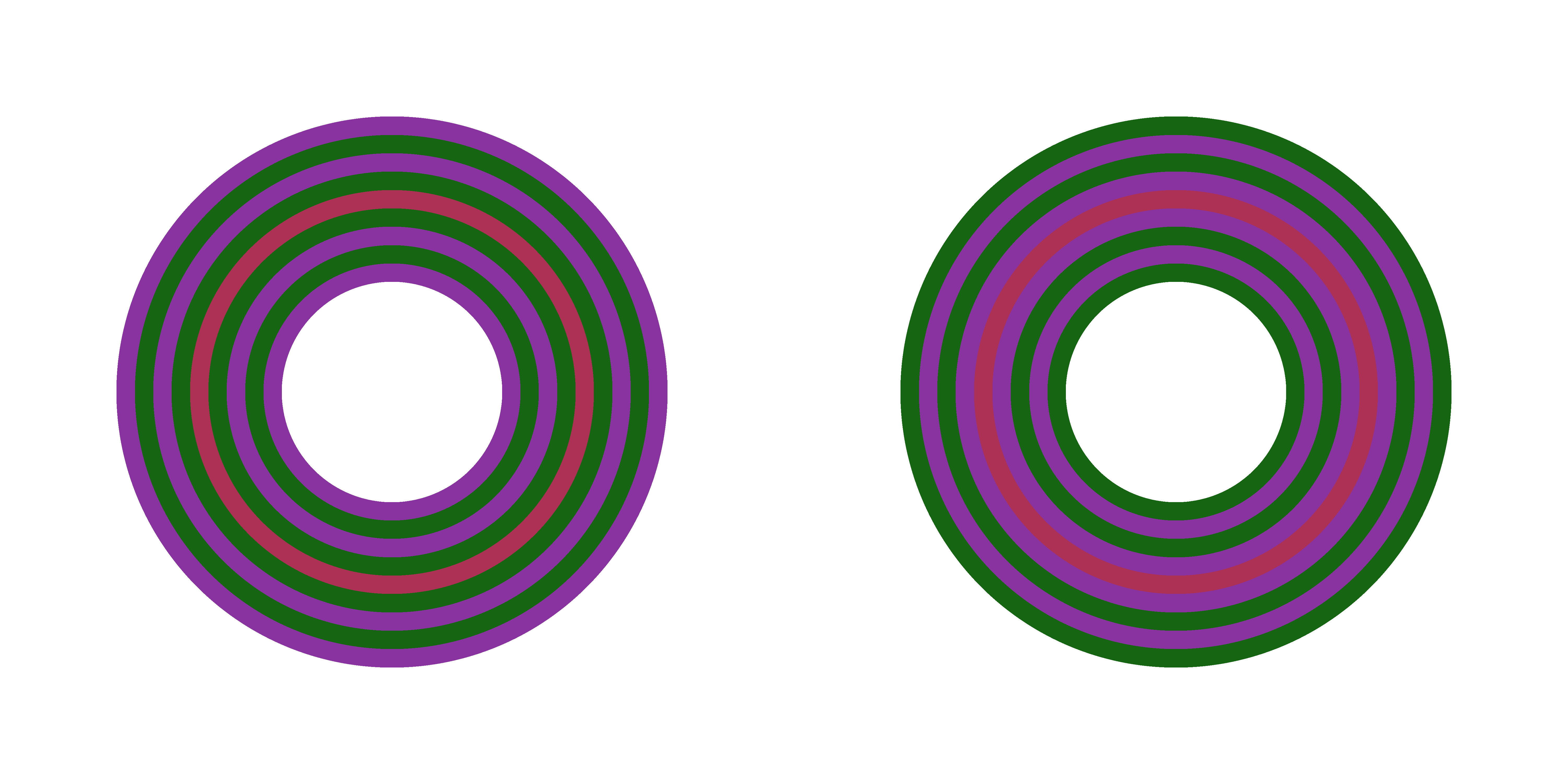} }%
    \qquad
    \caption{Induction examples. Left: lightness contrast; the center gray squares have the same luminance value but the one surrounded by white is perceived darker and the one surrounded by black is perceived lighter. Middle: lightness assimilation; all gray bars have the same luminance value but the gray bars surrounded by black are perceived as being darker than the ones surrounded by white, which are seen as being lighter. Right: chromatic induction; the central and inducing rings on both sides have the same RGB tristimulus values, but all rings are perceived differently due to their rearrangement.}
    \label{fig:indexp}%
\end{figure}

In visual perception, induction designates the effect by which the lightness and chroma of a stimulus are affected by its surroundings.
Visual induction can take two forms: assimilation, when 
the perception of an object shifts towards that of its surround, or contrast, when the appearance of an image region moves away from that of its local neighborhood. See Figure \ref{fig:indexp} for some examples.

\begin{figure}[!h]
\centering
\includegraphics[width=0.3\textwidth]{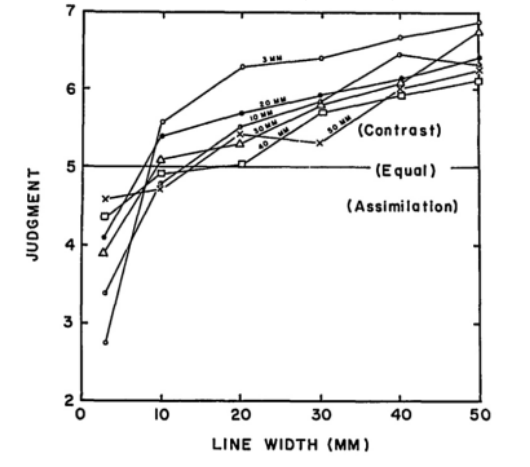}
\caption{Induction type depends on spatial frequency. Low spatial frequencies induce contrast, while high spatial frequences induce assimilation. Figure from \protect\cite{helson63}.}
\label{fig:helson}
\end{figure}

The groundbreaking experiments of Helson in 1963 \cite{helson63}
aimed to quantify the perceptual phenomena first formally described by von Bezold \cite{vonBezold74} and Gelb \cite{gelb30}, using matching experiments with printed induction bar patterns and isolated Munsell patches.
Specifically, observers had to judge the appearance of grey bars over white or black backgrounds.
When the bars were very thin, the observers reported assimilation; as the bars increased in width, the assimilation effect became less pronounced, and after some point the observers started to report contrast, whose effect became increasingly more pronounced as the width of the bars increased. See Figure \ref{fig:helson}. A similar result for the chromatic case was reported by Fach and Sharpe \cite{fach86}, who modulated the spatial frequency of patterns as opposed to the target background proportionality variation of Helson.
Their conclusion was that for higher spatial frequencies visual induction takes the form of assimilation, while for lower spatial frequencies it takes the form of contrast.
Though not all confirm these early observations, there exists a large body of later work (e.g. \cite{brenner03,brown97,harrar05,monnier03,shevell98,shevell05,wesner92}) corroborating the importance of the spatial distribution and variability of inducing surrounds.
Regarding visual induction models, we single out the work of Otazu et al. \cite{otazu10}, which is based on wavelet decompositions, and the very recent work of Song et al. \cite{Song2019}, that employs a neural field model.

In the film industry, the same movie is expected to be watched on displays of vastly different sizes, from cinema screens to mobile phones. But the typical viewing angle depends on screen size, being larger for larger displays, and therefore
the same image content will have a higher spatial frequency when seen on a small screen than when seen on a larger one.
As a consequence, the 
visual induction effects on both screens may not be of the same magnitude or even type: in the smaller display, induction effects of the contrast kind will have less magnitude and tend towards assimilation.

It is common practice in motion picture distribution to manually modify the original mastered picture when distributing to different display scenarios. In this process, a skilled artist works to ensure that the visual storytelling intentions of the piece in its original format are preserved. For instance, a piece which had an original theatrical release may be remastered for separate releases to home video, broadcast television, streaming, etc. However, new standards and developments in the display industry of the past decade (high dynamic range, wide color gamut, 4K and 8K displays, mobile devices) have increased the variability in potential content destinations to the point where manual processing is no longer a feasible solution for handling distribution masters. For this reason, it is an increasingly relevant effort for the motion picture industry to develop solutions to adjust content automatically considering the specific viewing scenario parameters of the viewer.

With the above mentioned in mind, in this work we make three main contributions. First, we show that a neural field model based on the efficient representation principle is able to predict induction effects, and this model is validated using existing psychophysical data. Second, we prove that by regularizing its associated energy functional the model becomes invertible. Finally, based on this invertible formulation we propose a method to pre-process an image in a screen-size dependent way so that its perception, in terms of visual induction, may remain constant across displays of different size. The potential of the method is demonstrated through novel psychophysical experiments on synthetic images and a validation experiment with natural images; all this data is made available as supplementary material.

\section{2. Neural Field Model for Induction Effects}\label{sec:eff}

The efficient representation principle, introduced by Attneave \cite{Attneave1954} and Barlow \cite{Barlow1961},
 is a general strategy observed across mammalian, amphibian and insect species, where visual processing considers the statistics of the visual stimulus and adapts to its changes \cite{Smirnakis1997}. In fact, efficient representation requires that the statistics of the image input are matched by the coding strategy, and while a global part of this coding strategy must have evolved on long timescales (development, evolution), in order to be truly efficient the coding must also adapt to the local spatio-temporal changes of natural images occurring at timescales of hours (e.g. from daybreak to dawn), seconds (e.g. when we move from one environment into another), or fractions of a second (e.g. when our eyes move around).
By constantly adapting to the statistical distribution of the stimulus, the visual system can encode signals that are less redundant and this in turn produces metabolic savings by having weaker responsiveness after adaptation, since action potentials are metabolically expensive \cite{Kohn2007}.

Atick \cite{Atick1992} makes the point that there are two different types of redundancy or inefficiency in an information system like the visual system:
\begin{enumerate}
\item
  {\bf  If some neural response levels are used more frequently than others.} For this type of redundancy, the optimal code is the one that performs {\it histogram equalization.}
  There is evidence that the retina is carrying out this type of operation at photo-receptor level \cite{Olshausen2000}, as their response curves match the cumulative histogram of the luminance distribution of the environment.
\item
  {\bf  If neural responses at different locations are not independent from one another.} For this type of redundancy the optimal code is the one that performs decorrelation. There is evidence in the retina, the LGN and the visual cortex that receptive fields act as optimal ``whitening filters'', {\it locally} decorrelating the signal.
\end{enumerate}

From the above, a local histogram equalization (LHE) process would simultaneously reduce both types of redundancy.
In \cite{Bertalmio:07}
Bertalm{\'\i}o et al. propose a variational method to improve the color appearance of images, that performs LHE.
They introduce the following energy functional, whose minimization yields the method's result:
\begin{eqnarray}\label{eq:vace}
  E(I)=\frac{\alpha}{2}\int_\Omega (I(x)-\frac{1}{2})^2dx - \gamma \int_{\Omega^2} w(x,y)|I(x)-I(y)|dxdy  + \nonumber \\ \frac{\beta}{2}\int_\Omega (I(x)-I_0(x))^2dx,
\end{eqnarray}
where
$I$ is an image channel in the range $[0,1]$, $\Omega$ is the image domain, $x,y$ are pixels,
$w$ is a distance function such that its value decreases as the distance between $x$ and $y$ increases, $I_0$ is the original image channel and $\alpha, \beta$ and $\gamma$ are positive weights.

The first term in the functional of Eq. \ref{eq:vace} measures the dispersion around the mid-range response of $\frac{1}{2}$, as in the {\it gray world} hypothesis for color constancy which states that in a sufficiently varied scene the average color will be perceived as gray (an observation made by Judd \cite{judd1940,judd79TU} and formalized by Buchsbaum \cite{Buchsbaum:80}) and therefore the illuminant color can be estimated from the color average of the scene; this implies that the minimization of $E(I)$ will make the image mean tend to $\frac{1}{2}$, so that the first term is small, and corresponding to the case where the illuminant is white.

The second term in the functional measures the contrast as the sum of the absolute value of the pixel differences (weighted, through $w$, by the distance between said pixels); because of the negative sign in front of this term, minimizing $E(I)$ will increase the contrast.

Finally, as the third term 
measures the difference with the original image $I_0$, the minimization of $E(I)$ will yield a result that can't be too far away from $I_0$.

The gradient descent equation for this functional is
\begin{equation}\label{eq:iter}
I_t(x)=-\alpha(I(x)-\frac{1}{2}) +\gamma\int_\Omega w(x,y)sgn(I(x)-I(y))dy -\beta(I(x)-I_0(x))
\end{equation}
Starting from $I=I_0$, equation \ref{eq:iter} is iterated until a steady state is reached (corresponding to a minimum of $E$), that will be the result of this algorithm.

The energy in Equation  \ref{eq:vace} introduces the influence of spatial neighbors through the distance function $w$. Without it (and with $\beta=0$) the energy becomes the one proposed by Sapiro and Caselles in \cite{Sapiro:97}, whose minimization produces a (global) histogram equalization of the original image. Therefore, we can argue that the evolution equation \ref{eq:iter} performs \emph{local} histogram equalization.

This method 
was applied channel-wise on color images in RGB \cite{Bertalmio:07} and in a color opponent color space like CIELAB \cite{Zamir2017}, and the results showed that the LHE method 
has several good properties:
\begin{enumerate}
\item 
  It has a very good local contrast enhancement performance, producing results without visual artifacts of any kind (only when the width of the locality kernel $w$ is very small do haloes start to appear).
\item
It ``flattens'' the histogram, approaching histogram equalization, as expected due to the relationship of Eq. \ref{eq:vace} with the one in the histogram equalization model of \cite{Sapiro:97}.
\item
  It reproduces visual perception phenomena such as simultaneous contrast and the Mach Band effect; this is consistent with the functional of Eq. \ref{eq:vace} modeling perceived contrast in a localized manner, with close neighbors exerting a higher influence than far-away points. 
\item
It yields very good color constancy results, being able to remove strong color casts and to deal with non-uniform illumination (a challenging scenario for most color constancy algorithms, as discussed in \cite{BertalmioBook}). 
\end{enumerate}

Additionally, the LHE model of \cite{Bertalmio:07}
is closely related to the neural field model of Wilson and Cowan, as pointed out in \cite{Bertalmio:07}
and further discussed in \cite{BC:09}. In particular, the evolution equation \ref{eq:iter} is very similar  to the Wilson-Cowan equations (see \cite{Bressloff:02,Wilson:72,Wilson:73}), which have a long and thriving history of modelling cortical low-level dynamics \cite{Cowan2016}.
It has been proven recently \cite{JNP2019} that the Wilson-Cowan equations are not variational, in the sense that they can't be minimizing an energy functional, and that the simplest modification that makes them variational yields the LHE method of \cite{Bertalmio:07}; furthermore, the LHE model provides a better reproduction of visual illusions than the Wilson-Cowan model.
The study of visual illusions has always been key in the vision science community,
as the mismatches between reality and perception provide insights that can be very useful to develop new models of visual perception \cite{Kingdom2011} or of neural activity \cite{Murray2013}, and also to validate existing ones.
It is commonly accepted that visual illusions arise due to neurobiological constraints \cite{Purves2008} that limit the ability of the visual system, and are therefore related to efficient representation.
In short, the LHE method (in its original formulation of \cite{Bertalmio:07} and also when it considers orientation \cite{JMIV2020})
is the generalization of the Wilson-Cowan equations that makes them compliant with the efficient representation principle, and 
at the same time this allows for an improved reproduction of visual perception phenomena.

\subsection{2.1 Modifying the LHE model so that it predicts induction}\label{sec:mod}
Looking at equation \ref{eq:iter}, we can see that the spatial arrangement of the image data is only taken into account by the weighting function $w$. But in practice $w$ is very wide, and therefore we can expect that the local contrast enhancement procedure of \cite{Bertalmio:07} will always produce contrast, not assimilation, since as we mentioned previously  assimilation is linked to high spatial frequencies \cite{shevell2003}.
In order to overcome the intrinsic limitations of \cite{Bertalmio:07} with respect to  induction, we should introduce spatial frequency in the energy functional. In \cite{BertalmioFrontiers2014} this is done by  making the parameter $\gamma$ in equation \ref{eq:iter}
change both spatially and with each iteration, according to the local standard deviation: if the neighborhood over which it is computed is sufficiently small, the standard deviation can provide a simple estimate of spatial frequency. But also, the standard deviation is commonly used in the vision literature as an estimate of local contrast.
The model in \cite{BertalmioFrontiers2014} can predict lightness assimilation and further improves efficiency by reducing redundancy: flattening the histogram and whitening the power spectrum.
Other attempts to modify the LHE formulation so that it better deals with induction are discussed in \cite{Bertalmio2019Book}.

Unfortunately, the modifications introduced to the LHE model in \cite{BertalmioFrontiers2014} do not fit well with the basic postulates of Wilson and Cowan's theory. This is why in this section we propose to adapt the LHE model in a different manner in order to predict induction, with changes that are motivated by neurophysiology data and that now keep the model consistent with the Wilson-Cowan formulation.
Specifically, we want to take into account the following biological phenomena.

\subsubsection{\it Photoreceptor response}
\begin{figure}[ht!]
\centering
  \includegraphics[width=0.4\textwidth]{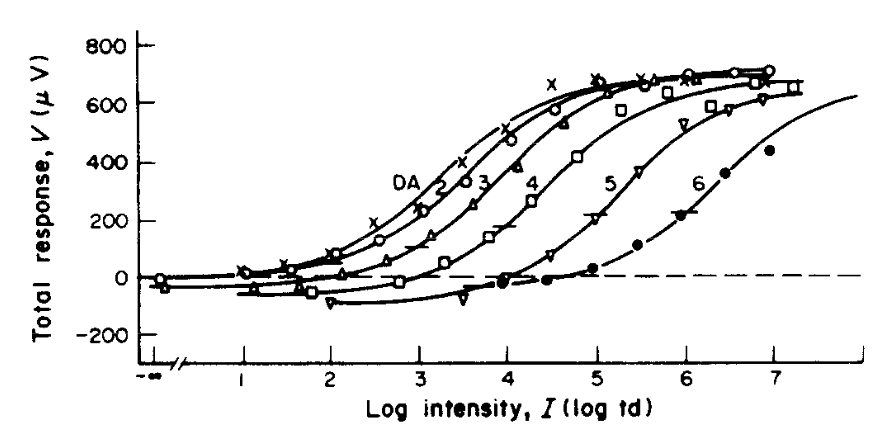}
\caption{Photoreceptor response curves for primate cones, for different background levels. Image from \protect\cite{Valeton1983}.}
\label{fig:vvn}
\end{figure}

  Photoreceptor response curves can be approximated very well with the Naka-Rushton equation:

\begin{equation}\label{eq:NR}
  R(I) = R_{max}\frac{I^n}{I^n+I_s^n},
\end{equation}

where $R$ is the response, $R_{max}$ is the maximum or saturation response, $I$ is the intensity, $n$ is an exponent of around $0.75$, and $I_s$ is the so-called semi-saturation value, the intensity at which the response is one-half of its maximum value and that roughly corresponds to the average intensity level. Notice that the Naka-Rushton equation is a monotonically increasing function and is therefore invertible; this point will become important later on.
If we increase $I_s$ and plot $R$ in linear-log coordinates, as in Figure \ref{fig:vvn}, then the curve moves to the right, the same curve-shifting phenomena observed when the background level increases. Therefore, light adaptation can be seen as changing the semi-saturation constant in the Naka-Rushton equation \cite{Shapley1984}. Furthermore, from Eq. \ref{eq:NR} and if $n=1$, we can obtain Weber's law. For this and other factors, it appears that the perceptual effects of light adaptation can be mostly accounted for by retinal processing \cite{Meister1999}.

\subsubsection*{\it Neural response nonlinearities and signal equalization}
\begin{figure}[h!]
  \centering
  \includegraphics[width=0.25\textwidth]{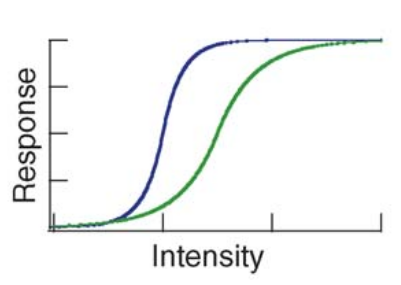}
  \includegraphics[width=0.25\textwidth]{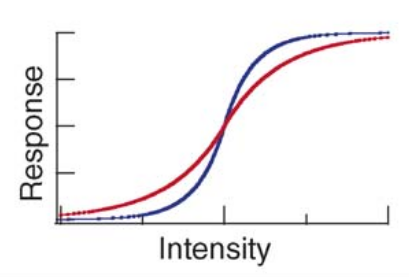}
  \caption{Neural adaptation to mean and variance. Left: neural response to higher (in green) and lower (in blue) mean luminance. Right: neural response to higher (in red) and lower (in blue) luminance variance. Adapted from \protect\cite{Dunn2006}.}
  \label{fig:dunn}
\end{figure}

Neural adaptation performs a (constrained) signal equalization by matching the system response to the stimulus mean and variance \cite{Dunn2006}, thus ensuring visual fidelity under a very wide range of lighting conditions. 
Figure \ref{fig:dunn} (left) shows that
when the mean light level is high, the nonlinear curve that models retinal response to light intensity is a sigmoid function with less steep slope than when the mean light level is low.
Figure \ref{fig:dunn} (right) shows that
at a given ambient level, the slope of the sigmoid is lower when the contrast is higher.
In both cases, the data is consistent with the nonlinearity of the neural response to light performing histogram equalization, since the nonlinearity behaves as the cumulative histogram (which is the classical tool used in image processing to equalize a histogram) does: darker images and images with lower contrast typically have less variance and therefore their cumulative histograms are steeper.
The psychophysical experiments in
\cite{Kane2016} corroborate that
the visual system performs histogram equalization
by showing how observers prefer display nonlinearities that allow the displayed image to be perceived as having a brightness distribution as close to uniform (i.e. with an equalized histogram) as possible.

\subsubsection*{\it Asymmetry of neural response nonlinearity}
\begin{figure}[h!]
    \centering
  \includegraphics[width=0.25\textwidth]{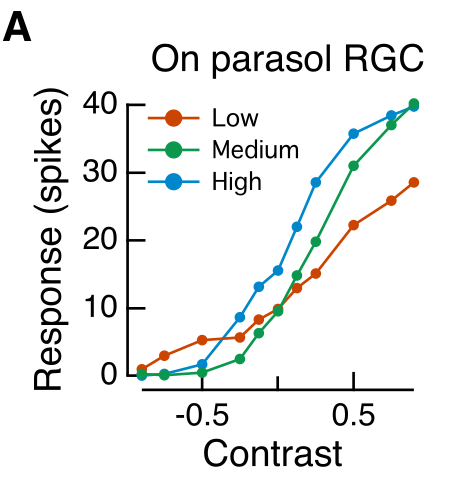}
  \includegraphics[width=0.25\textwidth]{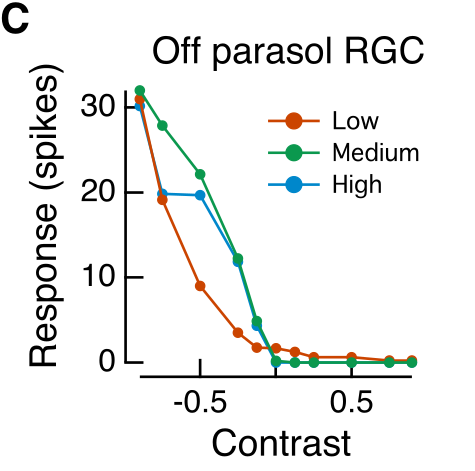}
  \caption{ON and OFF cells have different non-linear response functions. 
    Figure from \protect\cite{Turner2016}.}
  \label{fig:turner}
\end{figure}

Recent works from neurophysiology prove that OFF cells (those that respond to stimuli with values below the average stimulus level) change their gain more than ON cells during adaptation \cite{Ozuysal2012}, and that the nonlinear responses of retinal ON and OFF cells are different \cite{Kremkow2014,Turner2016,Turner2018},
see Figure \ref{fig:turner}.
This data on neural activity is consistent with psychophysical data \cite{Whittle1992,Kane2019} that demonstrates that 
our sensitivity to brightness is enhanced at values near the average or background level.

\subsubsection*{\it Retinal lateral inhibition can explain assimilation}
Lateral inhibition creates the typical center-surround structure of the receptive field (RF) of retinal ganglion cells (RGCs), with the excitatory center due to the feed-forward cells (photoreceptors and bipolar cells) and the inhibitory surround due to the inhibitory feedback from interneurons (horizontal and amacrine cells).
This center-surround organization is a very important instance of efficient representation, performing signal decorrelation and allowing to represent with less resources large uniform regions because they generate little or no activity.
It should also be pointed out that a more recent work \cite{Rucci2015} contends that decorrelation is already performed by the rapid eye movements that happen during fixations, and therefore that the signal arrives already decorrelated at the retina: the subsequent spatial filtering performed at the retina and downstream must have other purposes, like enhancing contrast.

Classical studies assumed that assimilation had to take place at a later stage than the retina, most probably at the cortex, because
it needs a much longer range of interaction between image regions than what lateral inhibition could provide with the classical RF size.
But in \cite{PLOS16} Yeonan-Kim and Bertalm{\'\i}o  showed that, in fact, assimilation can \emph{start} already in the retina. They took classic retinal models, those of Wilson \cite{Wilson1997} and van Hateren \cite{Van2005},
and adapted them so that parasol RGCs have a surround that is now dual, with a narrow component of large amplitude and a wide component of smaller amplitude. This different form for the surround is based on more recent neurophysiological data showing that
retinal interneurons have RFs that are much more extended than previously assumed, and RGC responses show a component that goes beyond the classical RF.

Based on the above we propose the following two-stage model:
\begin{enumerate}
\item
The image stimulus $I$, which is a scalar-valued linear image (i.e. an image channel proportional to light intensity) is passed through the photoreceptor nonlinearity, modeled as a Naka-Rushton equation, yielding $J_0$:
  \begin{equation}
   J_0= NR(I)=\frac{I^n}{I^n+{I_s}^n},
  \end{equation}
  where the exponent of the NR equation is chosen so as to maximize the equalization of the histogram of $J_0$, and the semi-saturation constant $I_s$ is the median average of the image.
\item
  The following evolution equation is run until a steady state is reached:
  \begin{eqnarray}\label{eq:iter2}
J_t(x)=-\alpha(J(x)-K_m*J(x)) +\gamma\int_\Omega K_c(x,y)\sigma(J(x)-J(y))dy \nonumber \\-\beta(J(x)-J_0(x)) 
  \end{eqnarray}
  Here $K_m,K_c$ denote kernels each expressed as a sum of two Gaussian functions and $*$ is the convolution operation, so now instead of a global mean $1/2$ as in Equation \ref{eq:iter} we have a local mean $K_m * J(x)$
  and local neighbors exert more influence but very far apart points can affect the response as well.  Furthermore $\sigma$ is a sigmoid function such that $\sigma(0)=0$ but not necessarily anti-symmetric, hence allowing positive and negative responses to be of different magnitude.
  Let us note that Eq. \ref{eq:iter2} is the gradient descent equation for an energy functional where the contrast term has this form:
  \begin{eqnarray}\label{eq:E2}
\int_{\Omega^2} K_c(x,y)\phi(J(x)-J(y))dxdy
\end{eqnarray}
where $\phi(\cdot)$ is a function whose derivative is the sigmoid $\sigma(\cdot)$.
\end{enumerate}
Let's call this model LHEI ($I$ for ``induction'') for the sake of brevity.

\subsection{2.2 Methods: LHEI Model Validation}\label{sec:exp1}

\begin{figure}[h!]
    \centering
    \includegraphics[width=0.3\textwidth]{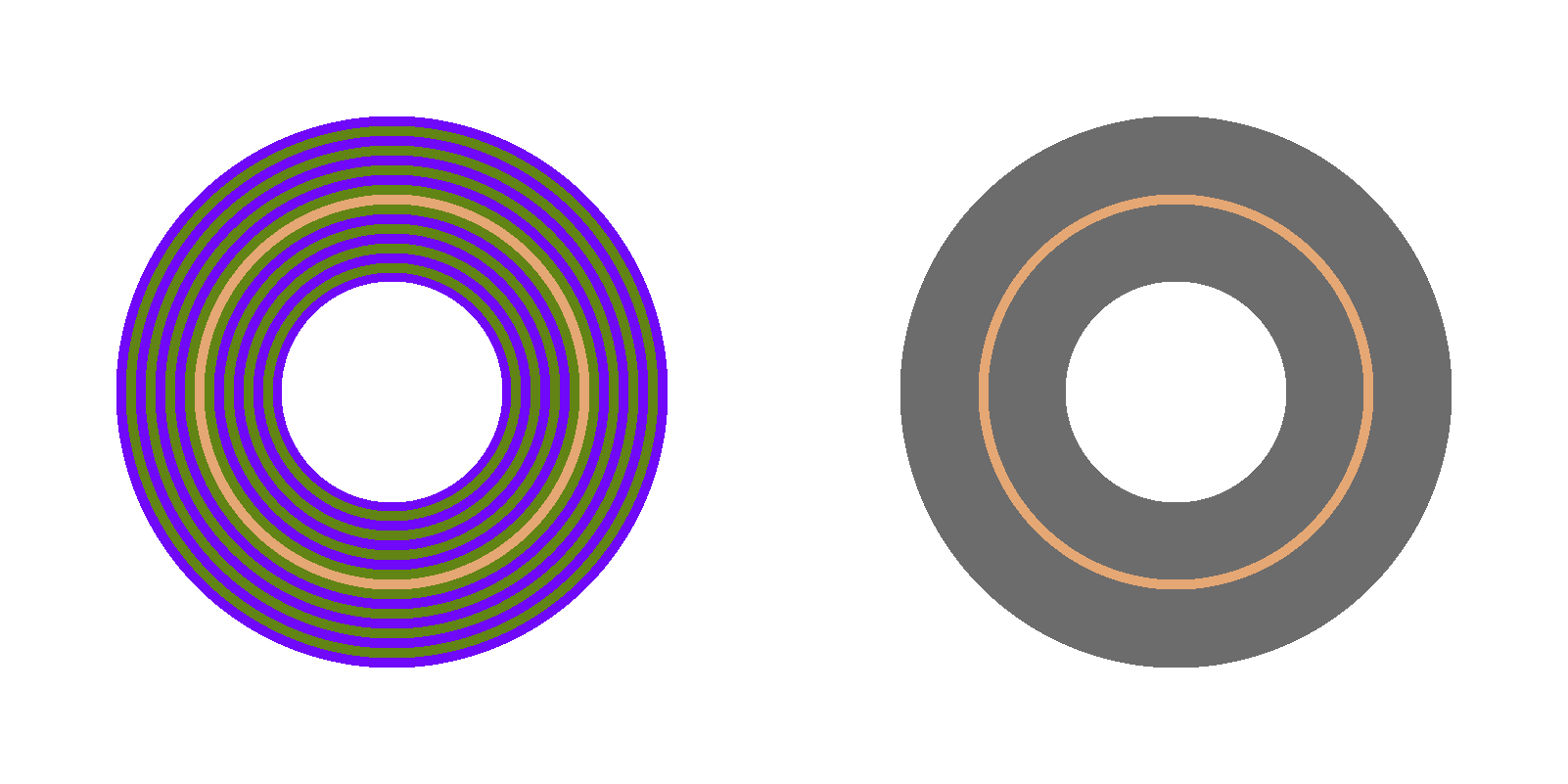} 
    \caption{Chromatic experiment stimuli. Left: test ring surrounded by concentric inducing rings of two alternating chromaticities. Right: comparison ring over uniform background. Note that the comparison and test rings are presented at the same chromaticity, and in the actual experiment, these patterns are placed over a black surround.}
    \label{fig:monnierstim}%
\end{figure}

In order to validate LHEI, we will use the chromatic induction data of \cite{monnier08}. In that work, observers were shown a test ring of some given chromaticity, surrounded by 16 concentric rings (half on each side of the test) that constitute the inducing pattern. This is the test image.
The surrounding rings alternated between two chromaticities, which in isolation appear lime and purple, selected because they differently stimulate the S cones only. Next to this image, the observer was shown a comparison ring, with the same dimensions as the test ring, but in this case simply presented over a uniform grey background (i.e. without inducing patterns). This is the comparison image.
Observers adjusted the hue, saturation and brightness of the comparison ring in order to match the appearance of the test ring.
\textcolor{black}{See Figure \ref{fig:monnierstim} for an illustration of this experimental set-up.}

The resulting chromaticity of the comparison ring is not the same as the chromaticity of the test, due to the induction effects produced by the lime and purple rings that surround the test ring: the difference in the S-chromaticity (associated to the S cones) between test and comparison rings is a color shift that quantifies the induction and can be plotted against the S-chromaticity of the test ring. Monnier performed this experiment with four observers, seven test-ring chromaticities, and the two possible alternating orders for the inducing rings (lime followed by purple, or the other way round). 

We have optimized the parameters of the LHEI model so that when we apply it to the test and comparison images, the resulting S-chromaticity difference between test and comparison ring is as close as possible to the one reported in the psychophysical experiments. For each of the initial conditions, we run our method using both the original rings, and the comparison ring adjusted by observers as input. Then, our minimization looks at the difference between the test ring in these two images. The error between the two images is computed as the $L_2$ difference between the value of the central rings. Finally, the error for each of the initial conditions is summed up to obtain the total error to minimize.

\subsubsection{Results}

\begin{figure}[h!]
    \centering
    \includegraphics[width=0.45\textwidth]{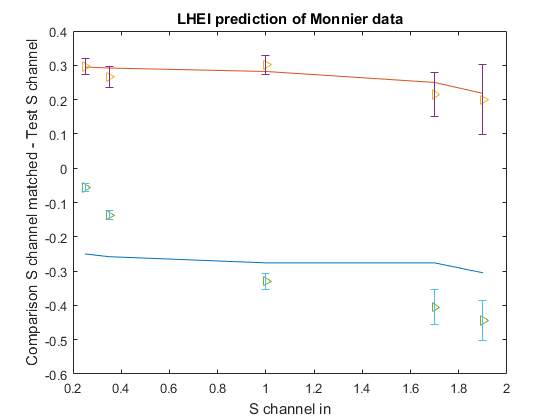} 
    \caption{Results from applying the LHEI model to observer data from \protect\cite{monnier08}. Triangles represent mean observer responses and lines represent LHEI model predictions, both in terms of S-chromaticity difference between test and comparison rings. 95\% confidence intervals are included for each of the observer data points. Values above zero: results when inducing rings next to the test ring have a lime hue. Values below zero: results when inducing rings next to the test ring have a purple hue.
    }
    \label{fig:songLHEI}%
\end{figure}

The resulting psychophysical data, averaged over the observers, is shown in Figure \ref{fig:songLHEI}
as orange triangles with purple error bars for the purple/lime patterns and green triangles with blue error bars for the lime/purple patterns. The error bars represent 95\% confidence error intervals about the mean, averaged across observers and trial repetitions. The fits of the model are shown in solid lines, in orange for the purple/lime pattern case and in blue for the lime/purple case. As we can see the fit is quite good, and qualitatively similar to the one obtained by \cite{Song2019} using a neural field model based on the Wilson-Cowan formulation. For the purple/lime pattern case, our model makes predictions that are within the range of experimental error for all test ring S channel values, however it does not properly fit the steeper slope of the lime/purple case.

\section{3. Invertible model for induction effects}\label{sec:explanationKernel}

As stated in the introduction, we want to derive a method that {\it matches} induction effects among screens of different size, not a method that {\it estimates} induction effects and their appearance. The difference is very relevant, and it's similar to the fact that colorimetry and color spaces allow us to determine quite accurately when two colors are perceived as different or the same, but they can't tell us the perceived appearance of said colors, as there are many external factors that play a role in this;  we must remark though that this approach contrasts with that of works like \cite{JNP2019}, where the output of the algorithm was explicitly simulating the appearance. 

Let's say that we have a color appearance model $M$ that is invertible and capable of reproducing induction effects. 
We consider two viewing scenarios A and B in which the same image stimulus $I$ is presented on a display, and both scenarios have identical viewing conditions except that the screen in A has a different size than the screen in B.
In this study we isolate for viewing angle and its effect on color perception; it is well known that other viewing parameters, like ambient illumination, screen luminance and dynamic range, display color gamut, etc. may have a significant impact on perception, but the usual practice in the literature, given the challenges in modeling vision, is to vary one of these elements while the others are kept fixed.
The model $M$ predicts for image $I$ an appearance $M_A(I)$ in scenario A and an appearance $M_B(I)$ in scenario B.
These appearances will be different because 
$M_A$ and $M_B$ are two instances of model $M$ that will generally have different parameter values. The reason for this is that, as we mentioned in section 2, neural processes adapt to the scene statistics, and by scene we mean the whole field of view, {\it a part of which} is the screen where the image stimulus is displayed. Therefore, different viewing angles will result in different scenes, consequently yielding different adaptation processes. 
In fact, in linear-nonlinear (L+NL) models of vision (and the model M we will be proposing shortly will be of this kind), adaptation is actually defined as the change of the model parameters when the input changes, and the full-view scenes in A and B provide different inputs to the visual system because the viewing angle of the screen is different.

Then, our induction matching goal can be expressed as determining the parameters for the compensation method $C= M_B^{-1} \cdot M_A$, because when the pre-processed image $C(I)$ is shown on screen $B$ its appearance, including induction effects, will be $M_B(C(I))=M_B \cdot M_B^{-1} \cdot M_A (I)=M_A(I)$, i.e. the same as if the image was seen on screen $A$.
In short, having an invertible appearance model $M$ for induction allows us to have an explicit analytical expression for $C$, and the parameter values for $C$ might be found so that they match psychophysical data. Furthermore, and very importantly, we don't need to optimize $M$ so that it accurately predicts induction effects in image appearance, which is a very challenging open problem: we just need to optimize $C$ so that the induction effects {\it match} in the two conditions.

This implies, however, that neither the LHEI model nor any of the color induction models in the literature (e.g. \cite{otazu10,Song2019}) can be used for our induction compensation goal, as they are not invertible.
In what follows we show how to modify the LHEI model so as to make it invertible.

In \cite{VSS16Kim} the authors went back to the retinal models that were updated and analyzed in \cite{PLOS2016}, studied what were their most essential elements, and produced the simplest possible form of equations to model the retinal feedback system that are nonetheless capable of predicting a number of significant contrast perception phenomena like brightness induction (assimilation and contrast) and the band-pass form of the contrast sensitivity function.
These equations form a system of partial differential equations that minimize an energy functional, closely related to the one of the LHE method of \cite{Bertalmio:07}, but where
 the absolute value function in the second term of Eq. \ref{eq:vace} is raised to the power of two.
This has the effect of \emph{regularizing} the functional, making it convex, and therefore its minimum can be computed with a single convolution, while the functional in \cite{Bertalmio:07} is non-convex and as a consequence its minimum has to be found by the iteration of the gradient descent equation.
If we follow this approach to modify the contrast term of the energy functional associated to the LHEI method (Eq. \ref{eq:E2}), we obtain:
  \begin{eqnarray}\label{eq:RE2}
\int_{\Omega^2} K_c(x,y)(J(x)-J(y))^2dxdy ,
\end{eqnarray}
where as usual $\Omega$ is the rectangular domain of the image that is displayed (i.e. not the whole field of view).

With this modification, the gradient descent equation previously shown in Eq. \ref{eq:iter2} now becomes:
    \begin{eqnarray}
J_t(x)=-\alpha(J(x)-K_m*J(x)) +\gamma\int_\Omega K_c(x,y)(J(x)-J(y))dy \nonumber \\-\beta(J(x)-J_0(x)) 
  \end{eqnarray}



Now the minimum can be computed directly by convolving the input image $J_0$ with a kernel $S$:
\begin{equation}\label{eq:S}
 S = \mathcal{F}^-1\left( \frac{\beta}{\alpha + \beta - \gamma -\alpha\mathcal{F}(K_m) + \gamma\mathcal{F}(K_c)} \right),
  \end{equation}
where $\mathcal{F}$ represents the Fourier transform. The kernel $S$ clearly has an inverse kernel $S^{-1}$ such that $S * S^{-1} = \delta$:
\begin{equation}\label{eq:Sinv}
 S^{-1} = \mathcal{F}^-1\left( \frac{\alpha + \beta - \gamma -\alpha\mathcal{F}(K_m) + \gamma\mathcal{F}(K_c)}{\beta} \right)
  \end{equation}

We propose the following modified version of the LHEI model, also consisting of two stages:
\begin{enumerate}
\item
  The first stage is identical to the first stage of the LHEI model:
  \begin{equation}\label{eq:M1}
    J= NR(I)=\frac{I^n}{I^n+{I_s}^n},
  \end{equation}
  where we recall that we consider $I$ to be a scalar-valued image, $I:\Omega \rightarrow [0,+\infty)$, so $J:\Omega \rightarrow [0,1)$.
\item
  The second stage produces the output $O$ as the convolution of $J$
  with the kernel $S$ of Eq. \ref{eq:S}:
  \begin{equation}\label{eq:M2}
    O = S * J
  \end{equation}
\end{enumerate}

Let's call this model $M$. 

The output $O=M(I)$ can be expressed as $M(I)=S*NR(I)$.
The inverse of the Naka-Rushton equation is
  \begin{equation}\label{eq:NR-1}
    NR^{-1}(J)=I_s\cdot(\frac{J}{1-J})^{\frac{1}{n}},
  \end{equation}
and the inverse kernel $S^{-1}$ was defined in Eq. \ref{eq:Sinv}.
Therefore, the inverse of $M$ can be expressed as $M^{-1}(O)=NR^{-1}(S^{-1}*O)$.


\section{4. Induction compensation method}\label{sec:ICM}

Based on model $M$, defined in Eqs. \ref{eq:M1} and \ref{eq:M2} above, we propose the following method for induction compensation for screens of different size.

If an image $I$ is to be shown on screen $B$ producing the same induction effects as if it were shown on screen $A$, in both cases under the same viewing conditions, then a compensation method $C$ must be applied to the image $I$, yielding an image $C(I)$.
When $C(I)$ is displayed on screen $B$ the induction effects are the same as when $I$ is displayed on screen $A$.
The compensation method $C$ is:
\begin{equation}\label{eq:C}
  C(I) = M_B^{-1} ( M_A (I) ) =  NR_B^{-1} ( S_B^{-1} * S_A * NR_A (I) )
\end{equation}

The linear filter $S$ of model $M$ has a center-surround form that, as mentioned in section 3, can perform decorrelation and contrast enhancement. For images with very high contrast, convolution with $S_A$ could produce over enhancement, resulting in some undershoot or overshoot values falling outside the range $[0,1)$, and in some cases these values might still remain out of range after convolution with $S_B^{-1}$, making it impossible to apply the function $NR_B^{-1}$ to them because its domain is $[0,1)$. To prevent these issues, in practice we clip all out-of-range values of $S_B^{-1} * S_A * NR_A (I)$ so that negative values are set to $0$ and values greater than $1$ are set to $1$; nonetheless, it is not expected that this clipping procedure produces visible artifacts, as attested by the natural image examples in Figure \ref{real_images}.

Following Equation \ref{eq:C}, our goal is to fit the two exponents of the Naka-Rushton equations and the two convolutions $S_B^{-1}$ and $(S_A)$. Following the approach used in section 3 to represent kernel $S$, we define what we call the compensation kernel $S_C$ as $S_C={S_B}^{-1} * S_A$:
\begin{equation}
S_C={S_B}^{-1} * S_A =\mathcal{F}^{-1}\left(\frac{D_2+D_1\mathcal{F}(K_F)}{C_2+C_1\mathcal{F}(K_F)}\right)
\end{equation}
where $K_F$ is the weighted sum of four Gaussians, and $C_1$, $C_2$, $D_1$, $D_2$ are real numbers.

The formulation above was presented for single-channel images. For color images, we will apply the induction compensation $C$ channel-wise. \textcolor{black}{To this end, given an input image in the display-referred RGB space, we will first transform it to a cone-space representation (CAT02 LMS space) \cite{CIECAM02} by applying the electro-optical transfer function (EOTF) of the input space, converting to CIEXYZ 2-degree tristimulus values given the chromaticity coordinates of the primaries and white point, and finally applying the 3x3 linear transformation matrix from XYZ to CAT02.} In this space, we will apply the first Naka-Rushton equation $NR_A$  to the individual L,M, and S channels. In this step the Naka-Rushton exponents will be equal for all the channels. 
After this is done, we will further convert the color representation to an opponent one, with channels that we call $Y$, $op_1$ and $op_2$, computed  as:
\begin{eqnarray}
Y = L+M+S\\
op_1 = L-M\\ 
op_2 = 2S-(L+M).
\end{eqnarray}
Then, our method will convolve each of the channels with the compensation kernel $S_C$. Let us note that there will be two different compensation kernels: one for the chromatic channels and another for the achromatic one. 
Once this is done, our method will apply the inverted opponent channel transformation, clipping to the range $[0,1)$  and the inverse of the second Naka-Rushton equation $NR_B$ (see Equation \ref{eq:C}). Again here, the Naka-Rushton exponent is kept equal for the three channels. \textcolor{black}{Finally, to convert the processed image back to a state that is ready for display, the inverse 3x3 linear transformation (CAT02 to XYZ) from the forward process is applied followed by the primary matrix (XYZ to RGB) and the inverse EOTF of the destination space.}

In order to validate our method we consider a scenario where $A$ corresponds to a cinema screen and $B$ to a mobile display.
We perform psychophysical experiments for both achromatic and chromatic induction patterns where observers look at a display with two scales of the same image, and they have to adjust the values of a given region of the small-scale image (corresponding to the mobile viewing scenario) so that it matches the appearance of that region on the large-scale image (corresponding to the cinema viewing condition).
For cinema, three picture heights viewing distance (a common figure for mastering) is assumed resulting in a vertical viewing angle of $18.92^{\circ}$; for the mobile condition, the same viewing angle as in \cite{canham18} is used resulting in a scaling factor of 0.39 between the two viewing scenarios. Using the data from these experiments the parameters of two separate kernels -one for the achromatic channel, another one for the chromatic channels- and the two Naka-Rushton exponents $n_A,n_B$ will be found by minimizing the error between the observer data and the method results.

\subsection{4.1 Methods: achromatic induction}\label{sec:exp2}

Following the preceding work \cite{bertalmio16}, the achromatic experiment was intended to be a direct expansion of the experiments of \cite{helson63} for the case of emissive stimuli. To this effect, we used the same type of induction pattern with a fixed inducer bar width and varied the comparison bar width. In this case however, observers reported the necessary correction factor directly by adjusting the luminance of the comparison bars in the mobile scaling to match those in the cinema scaling (test). The additional variable of starting comparison bar luminance was also varied between experimental presentations such that observers could approach their response from different directions. The complete matrix of experimental factors is shown in Table 1. 

\begin{table}[h!]
\centering
\caption{Achromatic experimental factors. Visual angles correspond to the cinema size patterns}
      \begin{tabular}{cccc}
        \hline
        Factor   & Type  & Levels \\ \hline
        Comparison width & variable & $0.19^{\circ},0.38^{\circ},0.54^{\circ},0.76^{\circ},0.96^{\circ}$\\
        Initial comparison luminance & variable & 4.0,8.1,22 $cd/m^2$\\
        Inducing bar width & constant & $0.19^{\circ}$\\ \hline
      \end{tabular}
\end{table}

\paragraph{Laboratory setup}
Experiments were conducted under dark surround viewing conditions on a calibrated Sony PVM-A250 reference monitor, representing an easily controllable cinema-like viewing environment. The monitor was calibrated to Rec. 709 primaries with a D65 white point and a 2.4 gamma decoding non-linearity. These settings were verified routinely before experimental sessions using a Klein K10-A colorimeter. The experimental cadence was controlled by a MATLAB test bed using the Psychophysics toolbox \cite{Psychotoolbox1,Psychotoolbox2} to display stimuli. Observers adjusted the comparison bar luminance via keyboard input to the experimental test bed.

\begin{figure}[h!]\centering
\includegraphics[width=0.45\textwidth]{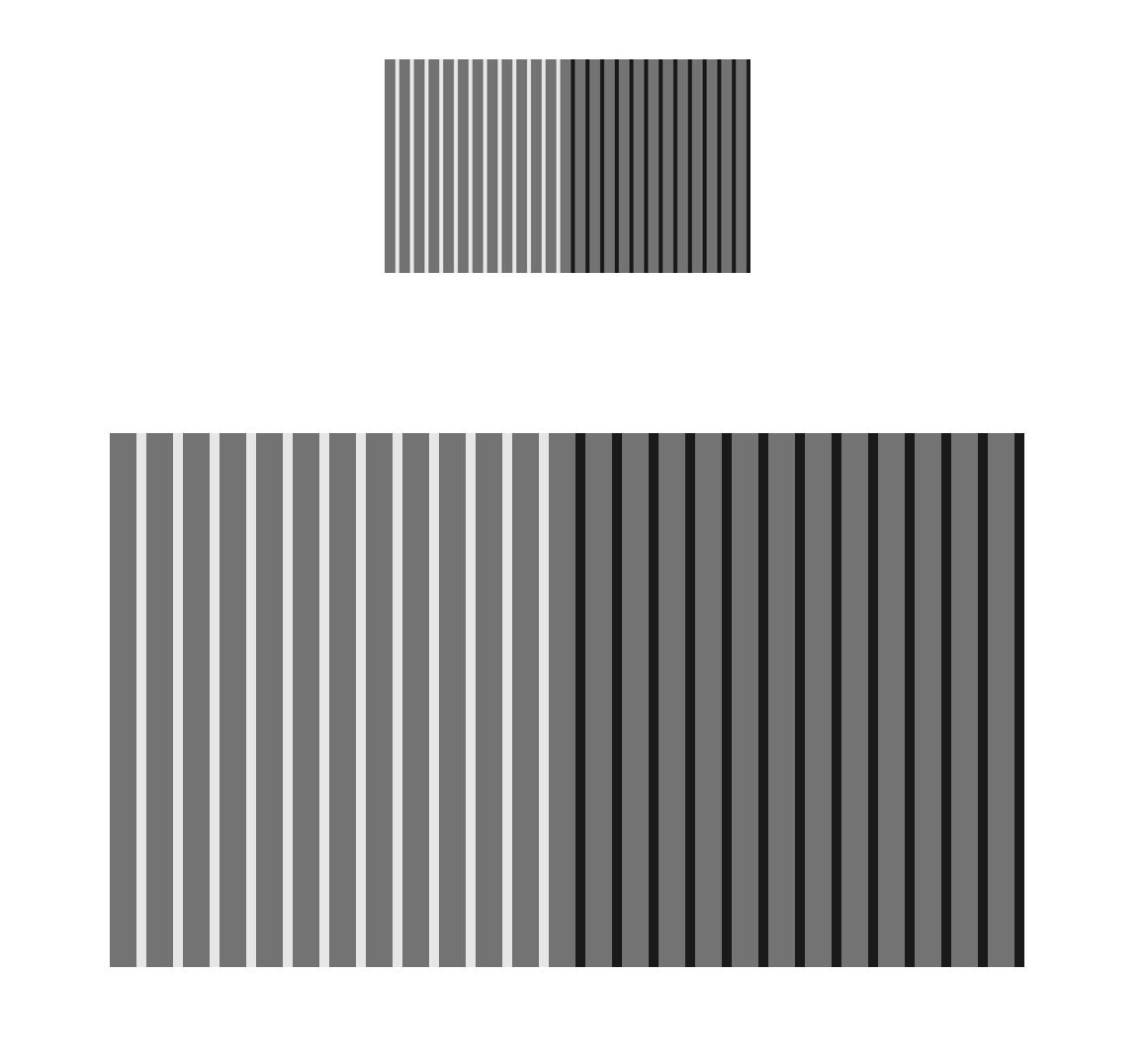}
\caption{Achromatic experiment stimulus. Note that all gray bars are presented at the same value, and in the actual experiment, these patterns are placed over a black surround.}
\label{Fig:astim}
\end{figure}

\paragraph{Stimuli}
Figure \ref{Fig:astim} shows the presented stimuli for the achromatic experiment. As can be seen in the figure, two patterns were presented on screen (except over a black background, as opposed to the white surround they are presented over in the figure.) Preliminary experiments showed that the background color was a relevant factor, so we elected to display patterns over a black background to match the dark surround viewing condition. Each pattern consists of two sides, representing positive (white/gray) and negative contrast (black/gray) respectively. Observers adjusted these two sides separately, but both are included simultaneously such that lightness references remain constant. The white fields were presented just below the maximum monitor white at a value of ~90 $cd/m^2$, while the black fields were presented at a value of ~0.6 $cd/m^2$.

\paragraph{Procedure}
Observers were ushered into the laboratory and the experimental instructions were read aloud. The instructions covered the purpose of the experiment, the observer task and the control scheme. Before starting the trials, observers were given a test trial with the experimenter in the room in order to familiarize themselves with the controls. Then, observers completed the experiment task for the 15 patterns with different target bar widths and presentation values as they were presented on the screen in a random order.

\paragraph{Observers}
Ten observers (2F,8M) aged between 23-39 took the experiment. All observers had normal or corrected acuity (20/20). Three observers are authors, while the remaining seven were naive to the purpose of the study.

\paragraph{Optimization}
The compensation model $C$ was optimized to fit the achromatic experiment data as generally as possible, meaning that the mean observer response from each target bar width experiment were considered simultaneously in the error function. The error was calculated as the sum squared difference in luminance values between the observer responses and a value sampled from the center of the comparison bars after the method was applied. By optimizing in this way, we show that the method can be made to work generally for different spatial configurations. 
The values obtained for this experiment following the above procedure are:
$n_A=0.7861$, $n_B=0.7063$, $K_F=-1.14G_{156}+1.86G_{29}+0.13G_{3}-1.76G_{40}$,  $C_1=3.94$, $C_2=2.54$, $D_1=2.46$, $D_2=2.72$. These values have been obtained in relation to images of size $800$ by $800$.

\subsubsection{Results}
\begin{figure}[h!]\centering
\includegraphics[width=0.45\textwidth]{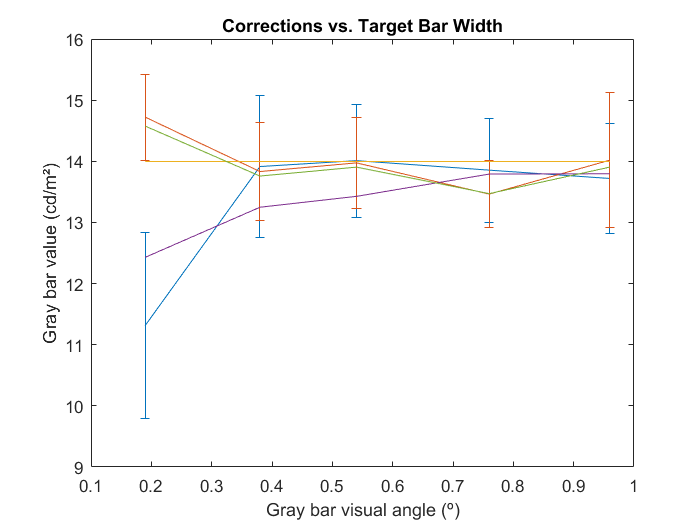}
\caption{Achromatic experiment results. Yellow: original gray value stimulus. Orange: average observer-selected value for gray bars over black background, showing $95\%$ confidence intervals. Blue: average observer-selected value for gray bars over white, showing $95\%$ confidence intervals. Green: prediction of induction compensation model $C$ for bars over black. Purple: model predictions for bars over white.}
\label{Fig:ares}
\end{figure}

Figure \ref{Fig:ares} shows the average observer responses in the experiment and the prediction provided by the compensation model $C$. We can see how the observer results are consistent with those  of \cite{helson63} in two key points: firstly, when the visual angle (equivalently the comparison bar line width) decreases, the appearance tends to assimilation, and hence the compensation requires enhancing the contrast;
secondly, as the visual angle increases, the amount of necessary compensation should decrease. Our model responses are consistently inside the range of experimental error for the observer data.

\subsection{4.2 Methods: chromatic induction}\label{sec:exp3}

Based on evidence that the phenomenon of induction occurs after visual signals are separated into different visual pathways, we expand on the achromatic experiments by making color matches in an opponent channel space, with the intention of applying our compensation model $C$ to the channels separately. In this case, we chose CIELAB space due to it having some degree of perceptual uniformity. In initial experiments we found that this expansion to three channel adjustment caused a great increase in the difficulty of the experimental task. Thus, we simplified the procedure by reducing the number of variables. In this case we test four color sets in the mobile and cinema sizes, and we test for three different comparison ring starting colors. To avoid observer fatigue, experiments were conducted two color sets at a time. The complete matrix of experimental factors is shown in Table 2.

\begin{figure}[h!]
    \centering
 \includegraphics[width=0.3\textwidth]{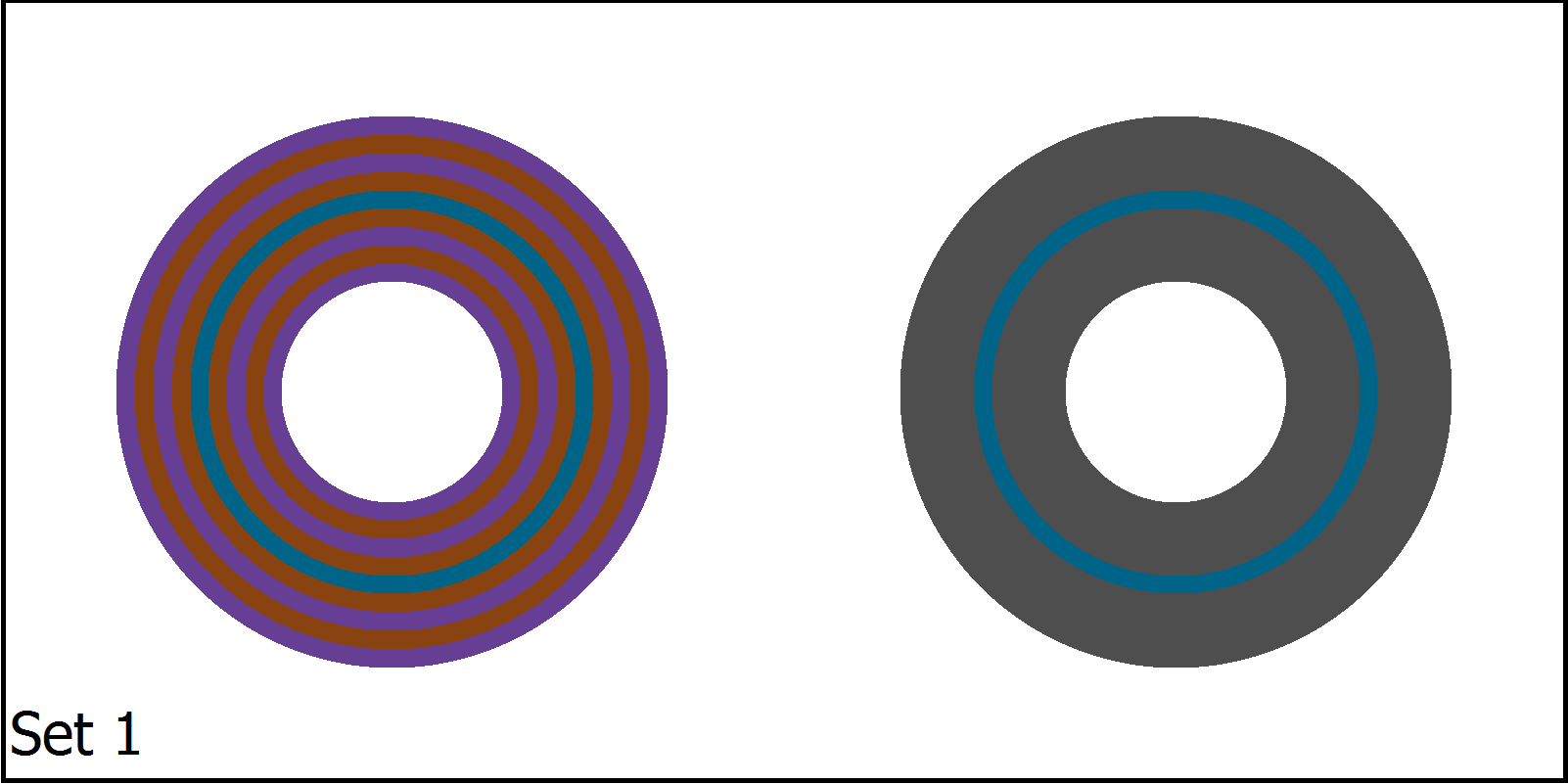} 
 \includegraphics[width=0.3\textwidth]{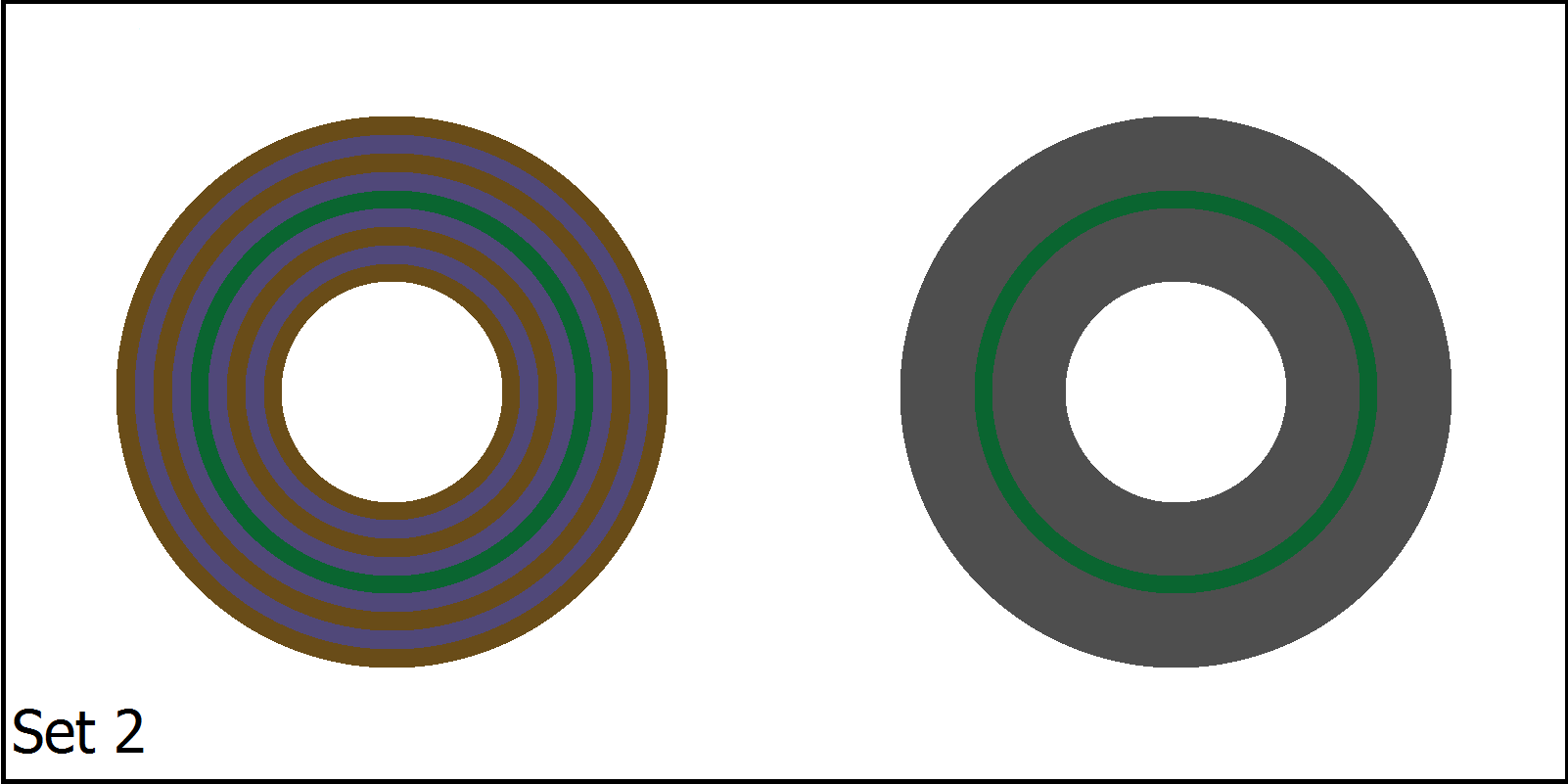}\\ 
 \includegraphics[width=0.3\textwidth]{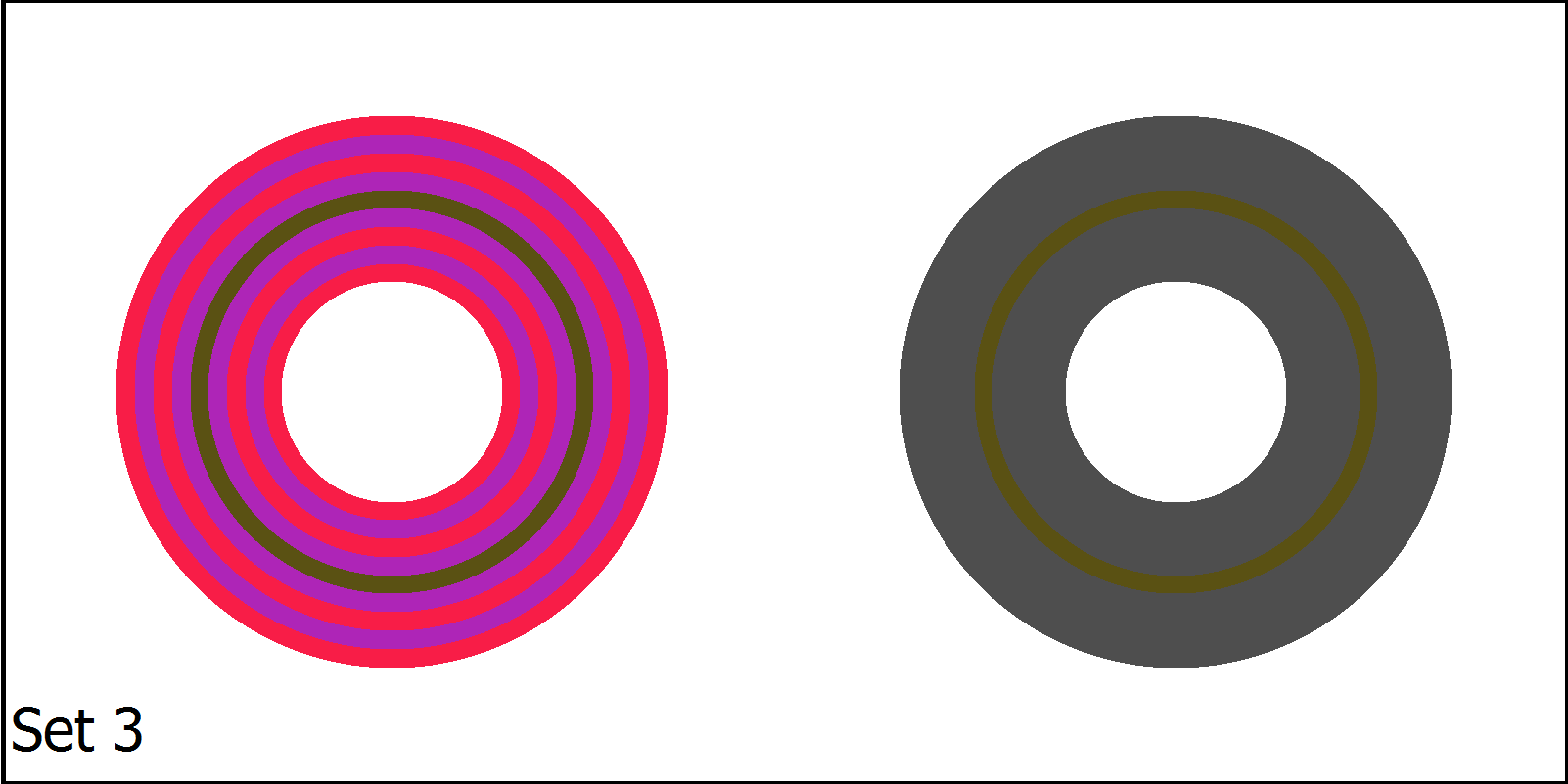} 
 \includegraphics[width=0.3\textwidth]{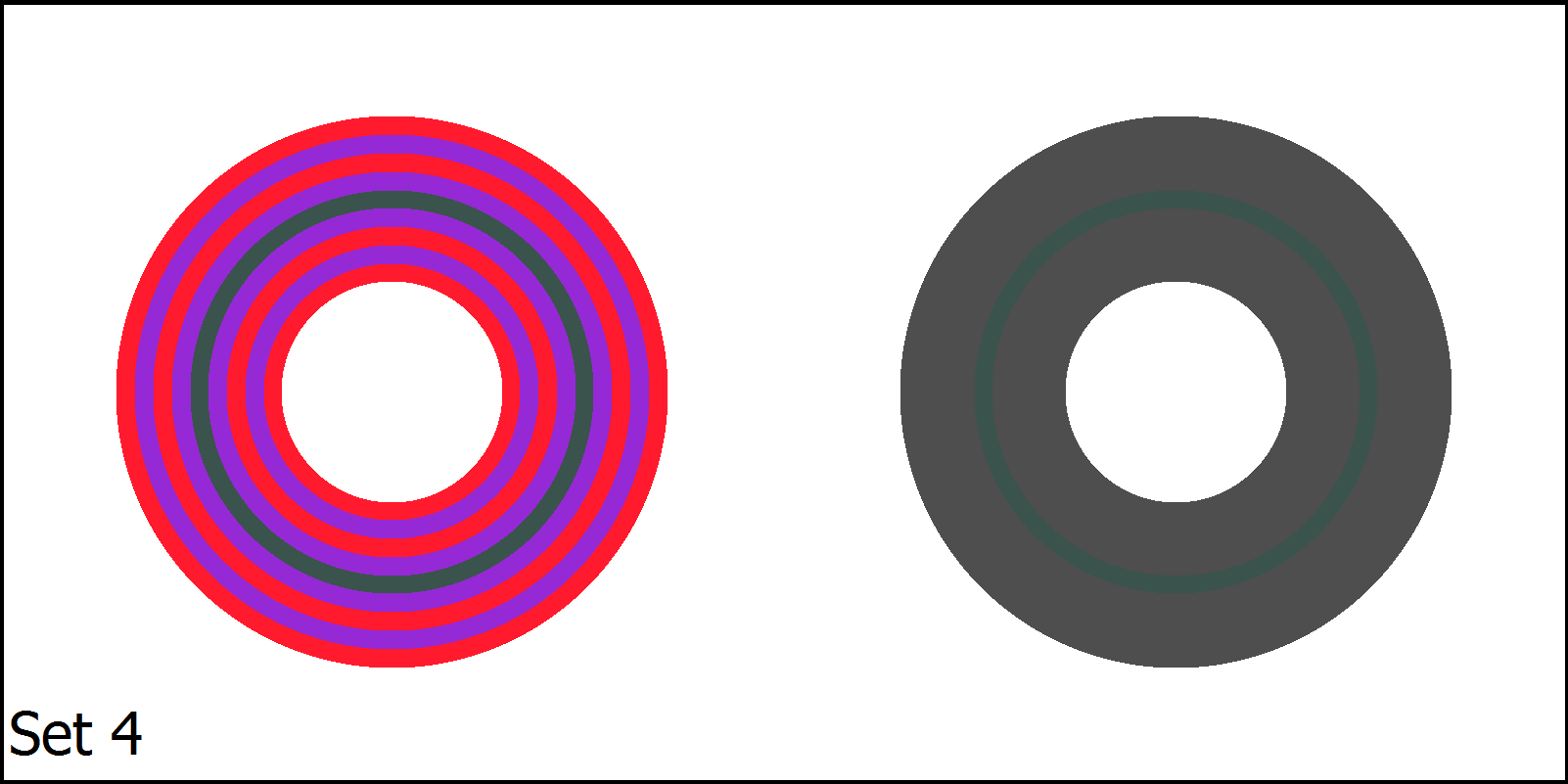}
\caption{Chromatic experiment stimuli corresponding to sets one through four. 
On the right side of each set is the comparison ring surrounded by an achromatic field, which observers were asked to match to the test ring on the left, surrounded by the induction pattern. To illustrate the strength of the visual illusion, the comparison and test rings are presented with the same RGB value here. Note that in the actual experiment, these patterns are placed over a black background.}
    \label{fig:chroStim}%
\end{figure}

For the chromatic experiments we took inspiration from \cite{monnier04} and used concentric circular induction patterns as shown in Figure \ref{fig:chroStim}. Observers must adjust the CIELAB values of the central ring of the comparison pattern (the achromatic circular ring on the right side of each set) so that it matches the appearance of the central ring of the test pattern (the concentric circular pattern on the left side of each set). This procedure was repeated for patterns at mobile and cinema scaling settings, and the observer reported correction was found by taking the difference between responses ($cinema - mobile$). Otherwise, an equivalent procedure to the achromatic experiment was conducted in this case.

\begin{table}[h!]
\centering
\caption{Chromatic experimental factors. Test pattern element sizes are relative to the cinema condition, but their size in proportion to each other is preserved for the mobile case.}
      \begin{tabular}{cccc}
        \hline
        Factor  & Type  & Levels \\ \hline
        Pattern scaling & variable & cinema and mobile (39\%) scaling\\
        Initial comparison level & variable & original, +10 a*b*, -10 a*b*\\
        Test pattern colors & variable & sets 1-4\\
        Test pattern diameter & constant & $11.0^{\circ}$\\
        Test to comparison distance (center to center) & constant & $15.6^{\circ}$\\
        Pattern center diameter & constant & $4.39^{\circ}$\\
        \hline

      \end{tabular}
\end{table}

\paragraph{Laboratory Setup}
The laboratory conditions regarding the display, surround, experimental test bed, were all the same as the achromatic experiment. One change, however, was that observers input their responses via a Tangent Element color correction panel which allowed for multi-channel adjustments to stimuli with separate knobs allowing for a more natural and reactive experimental interface in comparison to keyboard input.

\paragraph{Stimuli}
The stimuli for the chromatic experiments included four concentric circular induction patterns of different color arrangements similar to those used in \cite{monnier04}, shown in Figure \ref{fig:chroStim}. The two relevant features of these patterns are that their circular shape results in less after-images when compared to the bars, and their use of dual inducing colors leads to a stronger induction effect, allowing for more significant results to be gleaned from the experiment. The L* value of all rings in these patterns is kept consistent such that the focus of the observers' task could be on correction for chromatic induction. This said, observers were still permitted to adjust the L* channel value as equiluminance between pattern regions was not confirmed.

\textcolor{black}{In order to find patterns which exhibited a strong inductive effect, an experiment was performed in which 100 patterns containing regions with randomly selected L*a*b* values within the Rec. 709 color gamut were generated
(a color gamut is the range of colors that a display can reproduce, and Rec. 709 is the default gamut specification most commonly observed by display and television manufacturers).
These patterns were then shown side by side at the different scaling factors tested in the experiment. Then, patterns for which a hue shift could be identified between the different scaling factors were singled out. Finally, the experiment procedure was conducted for a single observer using all selected patterns from the previous step. From these results the final patterns were selected based on the criteria that the sensation of the target ring could be successfully reproduced in isolation, given the gamut of the monitor, and that a statistically significant correction (given 95\% confidence intervals) was called for by the observer between the two test pattern scaling factors. After several iterations of the experiment, six total color sets were found. Administering the test to multiple observers revealed that two of the sets should be removed, as the target colors were too close to the gamut boundary for observers to make reliable observations.}

\paragraph{Observers}
For the second color set, four observers participated in the experiment (1F, 3M) and for the remaining three color sets, three observers participated (3M). In both cases observer age ranged between 23 and 36, and two observers are authors while the remainder were naive to the purpose of the study. All observers had normal or corrected acuity (20/20).

\paragraph{Optimization}
 We performed these experiments for four different concentric ring patterns, then optimized our model $C$ so that it fits the data for three of these images and finally validated our results on the remaining image. To accomplish this, we apply the kernel $S_C$ only to the two chromatic components of our opponent channel space. \textcolor{black}{The forward model is otherwise applied as described earlier in the section, however during the inverse process we stop after applying the 3x3 transformation from CAT02 to XYZ, and convert this representation of the corrected image directly to CIELAB, taking the monitor white point of D65 at 100 $cd/m^2$ for the reference illuminant.} The optimization is performed in order to minimize the $\Delta E$ error on the test ring, and as we are using three different sets for training the minimization considers the maximum value of the $\Delta E$ error on the three test rings.
Since our method is working in a color opponent space different from CIELAB, when convolving kernel $S_C$ with our chromatic channels, shifts in the L* channel value may occur. To better comply with the observer responses, which reported no L* correction to be necessary, we decided to replace the L* channel of our result by the L* channel of the original image. 

The values obtained for the case where set 4 is used for testing (corresponding to results in Figure \ref{fig:chromaticResults} and column 5 in Table \ref{Table_training}) are: $n_A=0.5187$, $n_B=0.4439$, $K_F=-1.53G_{103}-0.67G_{43}+0.67G_{4}+0.34G_{26}$ , $C_1=2.81$, $C_2=1.30$, $D_1=2.27$, $D_2=1.60$. Let us note that these values have been obtained in relation to images of size $800$ by $800$.

\subsubsection{Results}
\begin{figure}[h!]
    \centering
 \includegraphics[width=0.35\textwidth]{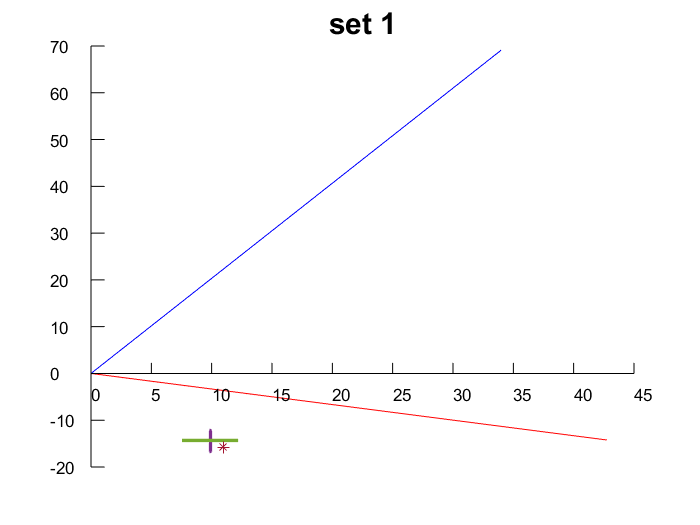}
 \includegraphics[width=0.35\textwidth]{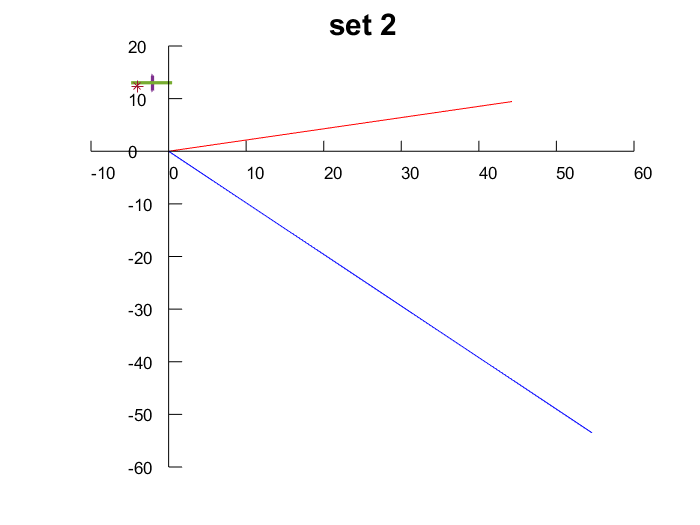}\\
 \includegraphics[width=0.35\textwidth]{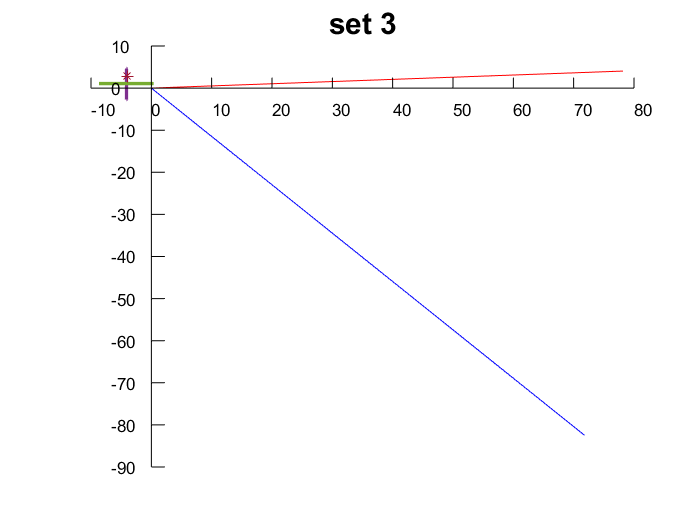}
 \includegraphics[width=0.35\textwidth]{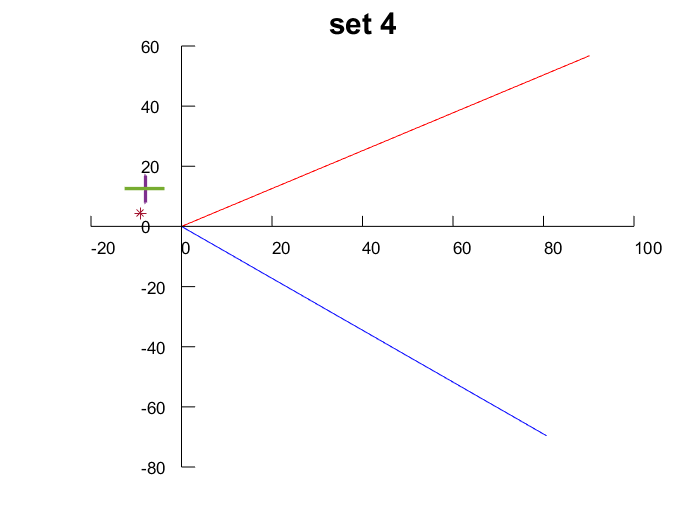}
\caption{Chromatic experiment results for the four tested color sets, with the a*b* value of the test ring centered at the origin. The vectors represent the magnitude and direction of color difference between the test ring and the inducing rings. The blue vectors represent the first inducers for each test case, or the color of the ring immediately adjacent to the test ring, while the red vectors represent the second inducer. The purple and green crosses represent the observer suggested compensation with 95\% confidence error bars, and the red stars represent the response predicted by our correction method. The bottom right plot (set four) was used to test the model fit while the remaining three were used to train the model parameters.}
    \label{fig:chromaticResults}%
\end{figure}

\textcolor{black}{In Figure \ref{fig:chromaticResults}, the results of the chromatic experiment are plotted in the two dimensional a*b* plane. The results are limited to the chroma channels, as the corrections reported by observers in the L* dimension were not statistically significant (95\% confidence error ranges overlapped the origin for all tested cases.) For each of the patterns the origin of the coordinate system is placed at the starting a*b* value of the test ring. In order to illustrate their directional influence, the plots depict the value of the inducing rings with a blue vector for the value of the inducing ring that is closer to the test ring, which we call the first inducer, and a red vector for the value of the other inducing rings, that we call the second inducer. The average observer response is depicted with purple and green 95\% confidence error bars.} 
  
\textcolor{black}{ Looking at the observers' responses, the induction compensation results selected by observers tend to show contrast mainly in the direction opposite to the first inducer, which implies that the appearance of the mobile viewing condition shows assimilation in the direction of the first inducer (because assimilation is compensated by contrast). In this way, the results for sets two through four were consistent with the classic assumptions on induction, as well as the results of \cite{helson63,fach86,monnier08}. However, our first set shows that this cannot be taken as a general rule, as observers reported the necessary correction to be roughly in the assimilation direction of the second inducer.}

\textcolor{black}{Regarding the ability of our compensation model $C$ for fitting this data, in Figure \ref{fig:chromaticResults} we added our results for the optimized values presented above. Our resulting corrections predicted by $C$ for each color set are depicted with a red star. We can see for the train cases that the predictions are within the range of experimental error, and for all cases compensate input in the proper direction.}

\textcolor{black}{To further study our model, Table 3 shows the error (measured as $\Delta E$ difference) between the average observer response and our model's prediction. As explained above, given that we have four different sets, we perform our experiments by training in 3 of the sets and testing in the remaining one. This gives us $4$ different cases. In the table, columns 2 to 5 represent each of the cases, with the model error for the testing set shown in blue; in particular, column 5 corresponds to the case illustrated in Figure \ref{fig:chromaticResults}. 
Column 6 presents in red what we call the 'original' error, 
the $\Delta E$ difference between the original data and the result of the observer correction. 
Finally, column 7 shows the improvement that our method presents over the original error, which is in the range $[45\%-60\%]$, therefore highlighting the advantage of applying our compensation method instead of doing nothing and just re-scaling the original image. 
}

  \begin{table}[]
  \centering
\begin{tabular}{c|cccc|c|c|}
\cline{2-7}
 & Train on: 2-4 & Train on: 1,3,4 & Train on: 1,2,4 & Train on: 1-3 & Original & Improvement \\
 & Test on: 1 & Test on: 2 & Test on: 3 & Test on: 4 & error &  (on test set) \\ \hline
\multicolumn{1}{|c|}{Set 1} & \textcolor{blue}{8,92} & 1,58 & 2,84 & 1,87 & \textcolor{red}{17,39} & 48,70\% \\ \cline{2-7} 
\multicolumn{1}{|c|}{Set 2} & 1,50 & \textcolor{blue}{5,35} & 2,84 & 1,95 & \textcolor{red}{13,20} & 59,47\% \\ \cline{2-7} 
\multicolumn{1}{|c|}{Set 3} & 1,61 & 1,58 & \textcolor{blue}{1,67} & 1,65 & \textcolor{red}{4,29} & 61,05\% \\ \cline{2-7} 
\multicolumn{1}{|c|}{Set 4} & 1,17 & 1,46 & 2,79 & \textcolor{blue}{8,18} & \textcolor{red}{15,00} & 45,48\% \\ \hline
\end{tabular}\label{Table_training}
\caption{Error between the average observer response and our model's prediction. We have performed the training for all combinations of 3 sets, testing on the remaining set (columns 2-5). Column 6 represents the original error, and column 7 represents our improvement, w.r.t. the original error in the test set.}
\end{table}

\subsection{4.3 Methods: induction in natural images}\label{sec:exp4}

Since the compensation method $C$ was designed with a direct imaging application in mind, it is important to analyze its effect in the context of natural images. In comparison to the synthetic stimuli used to optimize and validate the model, the context of real images introduces a significant increase in the spatial complexity of stimuli. For example, the image content could provide references for cognitive grouping feedback and other higher order processes which could be impactful to the induction effect \cite{murgia16}. Thus, before the method can be proposed for practical use, it is vitally important to first probe its behavior for a variety of test content and viewing contexts to ensure that it is tuned such that it improves the preservation of creative intent with changes in presentation size as a whole. 

In addition to the image content, our experiments revealed a number of additional viewing scenario factors which were influential to the induction effect. First, in initial iterations of the achromatic experiments, we found that the background and surround conditions can completely change the nature of the induction effect (changing the direction of required compensation). We also observed that induction effects are not only dependent on the relative scale adjustment between stimuli, but also on the absolute scaling. Due to this, a different correction would be required for stimuli with absolute scaling of 2 and 1, than for stimuli with absolute scaling of 1 and 0.5. Finally, our observers reported during the experiments that there were visible shifts in the appearance of the synthetic induction patterns with the amount of time spent viewing them.

\paragraph{Procedure}
With all of these factors in mind, an online validation experiment was designed and conducted to evaluate the performance of the method on natural images. The experiment was designed and distributed using the PychoJS library and the psychophysics-centered hosting platform Pavlovia \cite{pierce_2019}. In the experiment, observers were first given instructions to extinguish any direct light sources as to make their viewing environment as dark as possible. Then, observers conducted the virtual chin rest test of \cite{li2020} in order to determine their pixel-per-degree viewing angle so that stimuli could be adjusted to the correct presentation size. They were asked to maintain their seating position from this point in the experiment onward. Then, observers were given instructions which explained that they would be presented with original full-sized reference images and corresponding down-scaled pairs (some of which are altered with respect to their color and contrast and others which are unaltered) at timed intervals (5 seconds on, 2 seconds off), and would be asked to evaluate the match using one of the following options:

\begin{enumerate}
\item The colors of this image have been altered
\item The colors of this image may have been altered
\item The colors of this image have not been altered
\end{enumerate}

Following this, the observers conducted the body of the experiment, iterating through the test images in random order.

\paragraph{Stimuli}
The down-scaled versions include the original, unaltered image and the image corrected with our compensation method $C$. The method correction was applied following the process detailed at the beginning of the section, applying the kernel optimized in the achromatic experiment to the $Y$ channel and the kernel optimized in the chromatic experiment to the opponent channels $op_1$ and $op_2$. A series of images were selected which reflect a cinema or television shooting and grading style. The images cover a range of scene types and include important memory colors such as skin tones, product labels, natural colors, etc. To avoid observer fatigue, considering the two repetitions of each image and the minimum presentation time of approximately 10 seconds, the number of test images was limited to 33 to allow for observers to be able to complete the experiment with ample observation time in 20 minutes.

\paragraph{Observers}
A total of 16 observers participated in the experiment (10M/6F) aged between 24 and 58. None of the observers are authors and all were naive to the purpose of the experiment. Half of the observers work in an imaging related field and can thus be considered expert observers, while the other half were non-experts.

\subsubsection{Results}
\begin{figure}[ht!]
    \centering
    \begin{tabular}{cc}
    \includegraphics[width=0.23\textwidth]{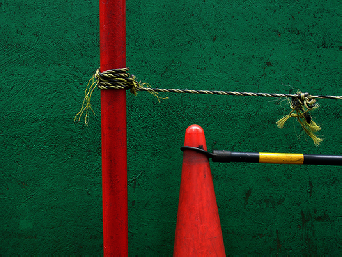} & 
    \includegraphics[width=0.23\textwidth]{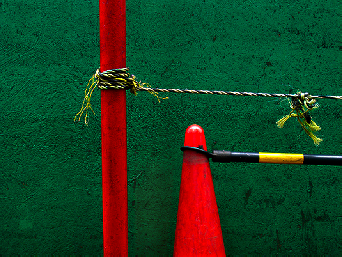}\\
   \includegraphics[width=0.23\textwidth]{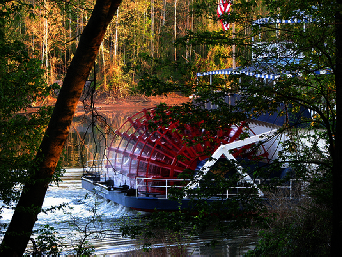} & 
    \includegraphics[width=0.23\textwidth]{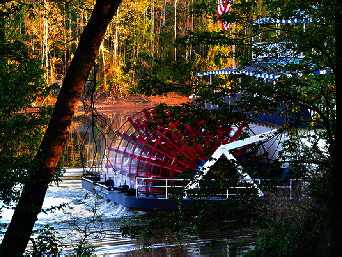}\\
   \includegraphics[width=0.23\textwidth]{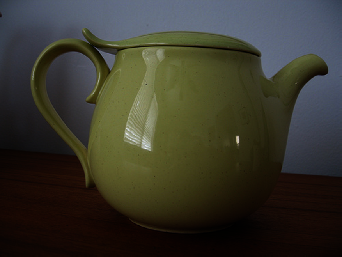} & 
    \includegraphics[width=0.23\textwidth]{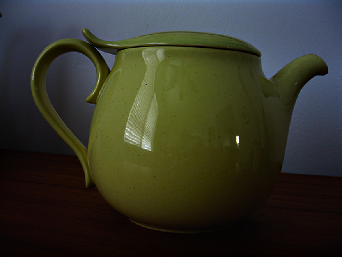}\\
       \includegraphics[width=0.23\textwidth]{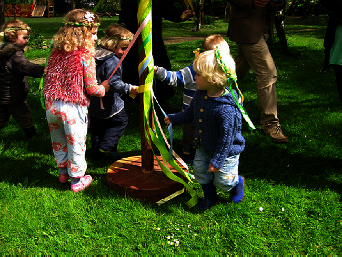} & 
    \includegraphics[width=0.23\textwidth]{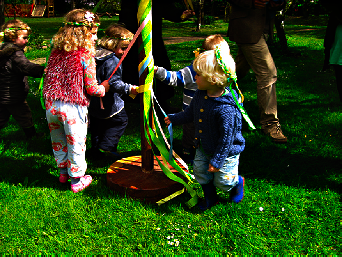}\\
          \includegraphics[width=0.23\textwidth]{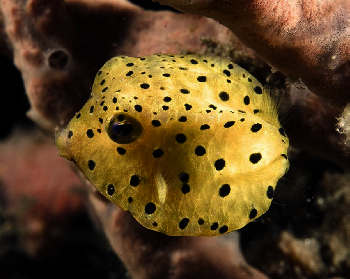} & 
    \includegraphics[width=0.23\textwidth]{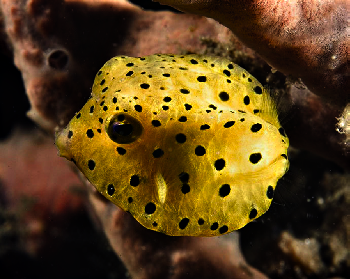}\\
        
   \end{tabular}
    \caption{Left: original Images. Right: results of our method with parameters optimized for training sets 1-3. Notice in our results how colors are slightly more vivid, and the absence of visual artifacts. Images sourced from \protect\cite{imagenet_cvpr09}}.
    \label{real_images}
\end{figure}

Figure \ref{real_images} illustrates the qualitative results of our induction compensation method $C$ on some natural images. We can see how in our results the colors are subtly but noticeably more vivid, e.g. the orange cone in the first row, the green teapot in the 3rd, the kid's blue jacket and boots and the grass in the fourth row, the yellow fish in the bottom row. This increased vividness corresponds to a contrast enhancement in the chroma, which should be cancelled out by the visual assimilation (and resulting chroma contrast reduction) produced when observing the image under a smaller field of view; the relationship between contrast enhancement and more vivid colors is discussed in detail in \cite{Zamir2017,Zamir2020,Bertalmio2019Book}.

While these results do not show visual artifacts of any kind, these problems can't be ruled out as they might appear if the method's parameters are optimized differently and/or the method is tested on other images.

\begin{figure}[h!]
    \centering
 \includegraphics[width=0.33\textwidth]{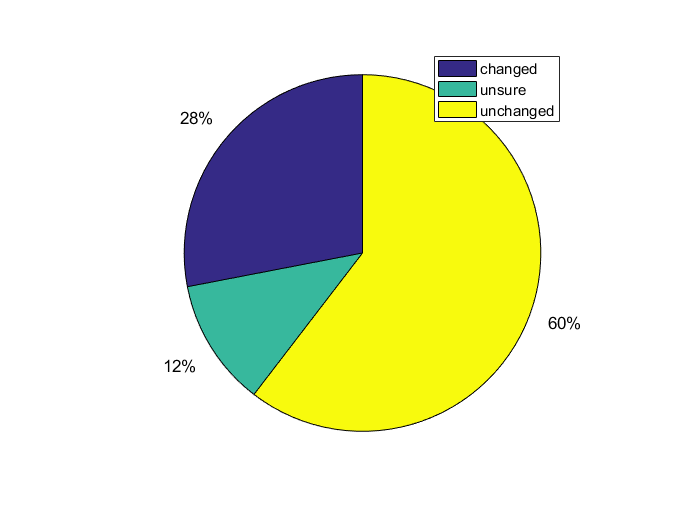} 
 \includegraphics[width=0.33\textwidth]{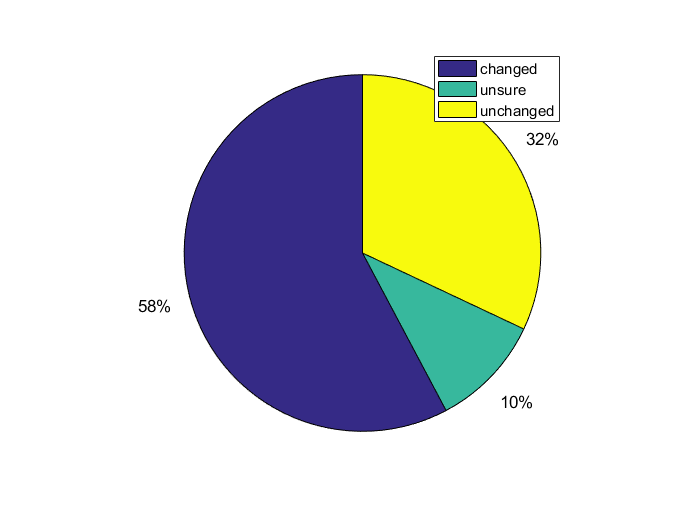}\\ 

\caption{Validation experiment results - percentages of each response for the control (left) and for our proposed method (right).}
    \label{fig:valPie}%
\end{figure}

In analyzing the quantitative experimental results, it is important to acknowledge that unlike the model optimization experiments, this experiment did not take place in a controlled laboratory setting and there could be significant variation in final image appearance between observers due to display performance and calibration, viewing environment limitations, and adherence to the experimental cadence. While we took steps to limit this variation in the online setting, this allowed for a lesser degree of control in comparison to the previous experiments. The results of Figure \ref{fig:valPie} show that, within this presentation context, the induction effect is subtle in natural images as in the majority of trials observers did not see a difference in color appearance between scaling settings. In contrast, the results of our method $C$ were seen as having had their color appearance shifted in the majority of trials. While this experiment was not a direct 2AFC comparison between the control and the results of our proposed method, these were the only two types of images presented to observers and thus the results can be interpreted comparatively, showing that the correction provided by our method was of greater magnitude than the shift caused by the induction effect in this scenario.

\section{5. Discussion}

The primary goal of these experiments was to determine the correction required to match the appearance of induction pattern targets at two different field of view scales. Based on the results of \cite{helson63,fach86}, we took the simple hypothesis that a greater degree of contrast would always be observed in the larger field of view pattern. Thus, the correction from small pattern to large for a given channel should always be in the direction of contrast, and the results of the achromatic experiment confirmed our hypothesis.

 \textcolor{black}{For the chromatic experiments, making the assumption that the phenomenon of induction occurs after visual signals are separated into different visual pathways, we chose to make color matches in an opponent space with the intention of applying our corrective method $C$ to the channels separately, with different optimization values compared to the achromatic case. We found that the use of the simple bar patterns of \cite{fach86} caused a multitude of problems in the chromatic case. Observers reported weak induction effects as well as strong afterimages when shifting their gaze between test patterns. In addition, the direct comparison of the cinema-sized pattern to the mobile pattern was confusing to observers, as the inducers in the smaller pattern appeared to be significantly less saturated. As a solution, we took inspiration from \cite{monnier04} and used concentric circular induction patterns, as shown in Figure \ref{fig:chroStim}, whose main features are that their circular shape results in less afterimages when compared to the bars, and their use of dual inducing colors leads to a stronger effect, allowing for more significant results to be gleaned from the experiment.}
 
\textcolor{black}{A procedure for the selection of color sets which exhibited a strong induction effect is explained in section 4.2. While this experiment was more or less informal in nature, its results demonstrate the rarity of strong induction effects given random color combinations, even if they are arranged in synthetic patterns which emphasize induction. Another interesting anomaly which can be observed from this experiment is that despite the random nature in which they were generated and selected, the final four patterns appear quite similar to each other, all containing an inducer of violet hue.}

The results of the chromatic experiments presented here clearly show that the original hypothesis (that contrast effects will shift towards assimilation with an increase in test pattern spatial frequency) can be broken. While three of the color sets showed this behavior, we can see that the first color set breaks the trend, and is closer to requiring correction in the assimilation direction with respect to the second inducer. One clue to this differing behavior is related to the violet inducers which appear in each pattern. In all patterns for which induction effects behaved as expected, the violet inducer was directly adjacent to the test and was the primary induction influence. However, for the first color set, the violet field serves as the second inducer which is not directly adjacent to the test field, but still acts as the primary induction influence.

Outside of this data, we also found patterns which broke our simple hypothesis in preliminary experiment iterations. From these iterations we observed that the luminance level of the background/surround and the hues of inducing and target patches to be relevant factors. This type of conflicting and paradoxical finding seems to be common in the study of induction, with many works being based on the discovery of scenarios which contradict previous findings \cite{monnier08, murgia16}. While this phenomenon can be found in all research topics and is a sign of progress, its frequency in this area is an indication that induction as a whole is still very much an open problem, despite its earliest formal works dating back nearly a century and a half. This can be justified by the fact that the phenomenon is the result of complex interactions involving both physiological and cognitive processes \cite{singer94} on multiple visual pathways.

While the model $C$ was designed with the intention that it would always correct the test targets in the contrast direction, we have had some success in optimizing a kernel which works more generally in fitting to these four test cases. These results were garnered by model fitting using three chromatic sets and testing on a fourth one. Looking at Table 3, one can see that the model $C$ performs slightly better in predicting examples which are within its test set in comparison to when they are excluded, implying that its behavior is somewhat biased towards its training set. 

While the achromatic training set included a number of cases for which the necessary induction correction was negligible, our chromatic kernel is trained exclusively on examples which required a large correction between presentation sizes. Our process of selecting these examples showed that this effect is quite rare, even among synthetic patterns which are specifically designed to produce strong induction effects. The results of the validation experiment showed that the effect is even more subtle in the case of natural images when presented in a "wild" context with variation in display and viewing environment conditions, with a significant majority of observers reporting no color shift in the control examples. In comparison, the correction provided by our method was detectable by observers in the majority of cases. Thus, it is likely that our method in its current optimized state is producing an exaggerated correction for what the practical application requires. For this reason, the compensation method presented here is to be interpreted as a proof of concept 
and a contribution to the research in the field
as opposed to a procedure which is ready to be used in practice. 

A further interesting challenge is that the compensation value observers reported to adjust between mobile to cinema appearance could be outside of any given monitor gamut space, or outside of the gamut of physically realizable colors, depending on the position of the test target and the magnitude and direction of the induction shift between screen sizes. In these scenarios, induction effects will only be partially compensated for by the method $C$. We encountered this issue with three of our four test sets, and opted to clip all observer and kernel reported corrections to the Rec. 709 gamut. By doing this, our model's results can be readily reproduced by the most common displays, including those which we used for visual proofing during its development.
 \textcolor{black}{We later performed a preliminary analysis with input encoded under the larger standard color gamuts Rec. 2020 and CIE 1931 XYZ, where the clipping of observer corrections is smaller for the former and almost negligible for the latter. The results showed that the method $C$ makes corrections of similar accuracy when it is required to reach out into larger color volumes.}

\section{6. Conclusion}

In this work we have shown that a neural field model performing local histogram equalization is able to predict chromatic induction effects. This is a variational model, an embodiment of the efficient representation principle, and by regularizing its associated energy functional the model is still able to represent induction and now  becomes invertible. This fact allows us to use the new invertible model as the basis for an induction compensation method, which we call $C$, to pre-process an image in a screen-size dependent way so that its perception, in terms of visual induction, may remain constant across displays of different size. The potential of the method is demonstrated through psychophysical experiments on synthetic images, both achromatic and chromatic. Our results show that the established assumption in the literature that induction tends towards assimilation as the spatial frequency increases is sometimes contradicted by the experimental data, and therefore can't be taken as a general principle.

We believe there are three main avenues to explore in order to improve our proposed approach:
\begin{enumerate}
\item
  Our induction compensation technique $C$ is based on a color appearance model that follows the classic formulation of a cascade of linear-nonlinear (L+NL) modules \cite{Martinez2018} and has a biological correlate, consisting of a nonlinear stage (the Naka-Rushton equation that models photoreceptor responses) followed by a linear stage (convolution with a kernel that models lateral inhibition in the retina).
  A L+NL model is valid for stimuli of a given distribution  seen under given viewing conditions, in which case it may provide a good match to the firing rate.
  But visual adaptation, an essential feature of the neural systems of all species by which changes in the stimuli produce a change in the input-output relation of the system  \cite{Wark2009}, alters the visual system response. Visual adaptation is clearly a key element of the efficient representation principle and it affects, among other things, the spatial receptive field and temporal integration properties of neurons, requiring changes in the linear and/or the nonlinear stages of a L+NL model in order to explain neural responses \cite{Meister1999}. So, for example, depending on the input the receptive field of a single neuron can have different sizes or preferred orientations \cite{Coen2012}, or even change polarity (ON/OFF) \cite{Jansen2018}.
For our purposes of induction compensation, we should study how to make the convolution kernel $S$ depend on the input, or instead to use a filter bank as is the traditional approach with L+NL models for visual perception \cite{Wandell1995,Graham2011}.
\item
  Another option is to study how to make the LHEI model invertible while keeping it as a nonlinear neural field model (i.e.  without regularizing its associated functional), looking into a gradient {\it ascent} equation or alternatively considering changing the sign of the parameter $\gamma$ in the model, as it has been shown that with one sign for $\gamma$ the model increases the contrast while with the opposite sign the model reduces the contrast \cite{Bertalmio:09,Zamir2020,Zamir2014}.
  We believe this option has more potential because the resulting compensation model would not be of L+NL form.
\item
  Finally, a third avenue to explore, compatible with the previous two, would be to design and carry out psychophysical experiments for induction compensation where observers are asked to adjust values over the whole image and not just on a particular region like the gray bars or the test ring. \textcolor{black}{Using this data, the correction method $C$ could be optimized such that it produces a balanced correction for all of the spatially adjacent regions in the patterns simultaneously, accounting for their interdependent effects. Additionally, it would be an interesting expansion of the work to include more test sets in the chromatic case which do not produce strong illusions, as was done in the achromatic case, as this would likely better represent the behavior of the visual phenomenon in response to natural images.}
\end{enumerate}

\section*{Acknowledgements}
The authors would like to thank those who served as experimental observers, who this work would not have been possible without. In addition, we would like to thank Patrick Monnier and Anna Song for their help in obtaining the data from \cite{monnier08}. This work has received funding from the European Union’s Horizon 2020 research and innovation programme under grant agreement number 761544 (project HDR4EU) and under grant agreement number 780470 (project SAUCE), and by the Spanish government and FEDER Fund, grant ref. PGC2018-099651-B-I00 (MCIU/AEI/FEDER, UE). JVC received funding from the project PID2019-109628RJ-I00/AEI/10.13039/501100011033 by the Ministerio de Ciencia, Innovación y Universidades (MCIU) and the Agencia Estatal de Investigación (AEI) of the Spanish government.

\section*{Abbreviations}
LGN - lateral geniculate nucleus \\
LUT - look up table \\
PDE - partial differential equation \\
LHE - local histogram equalization \\
RF - receptive fields \\
RGC - retinal ganglion cells \\
DoG - difference of gaussians \\
L+NL - linear-nonlinear
2AFC - Two alternate forced choice

\end{document}